\documentclass[a4paper,11pt]{article}

\usepackage{amsmath,amsfonts,amssymb,theorem,verbatim,textcomp}
\usepackage{amsfonts,amssymb,graphics,theorem}
\usepackage[greek, english]{babel}
\usepackage[pdftex]{graphicx}

\usepackage{float,multirow}
\floatstyle{ruled}
\newfloat{algo}{thp}{lop}
\floatname{algo}{Algorithm}

\newtheorem{Thm}{Theorem}
\newtheorem{Cor}{Corollary}[section] \newtheorem{Lemma}{Lemma}[section]
\newtheorem{Prop}{Proposition}[section] 
{\theorembodyfont{\normalfont}

}

\def\BEN{\begin{enumerate}}  \def\BI{\begin{itemize}}
\def\EEN{\end{enumerate}}   \def\EI{\end{itemize}} \def\im{\item}
   
  \def\no{\noindent}

\def\nn{\nonumber}
\def\beq{\begin{eqnarray}} \def\eeq{\end{eqnarray}}

\def\al*#1{\begin{align*}#1\end{align*}}

\def\ga*#1{\begin{gather*}#1\end{gather*}}

\def\alat*#1#2{\begin{alignat*}{#1}#2\end{alignat*}}
\def\bea{\begin{eqnarray*}}
\def\eea{\end{eqnarray*}}
\def\ml*#1{\begin{multline*}#1\end{multline*}}

 \def\mbf{\mathbf} \def\mrm{\mathrm}
 \def\unl{\underline} \def\ovl{\overline}

  \def\R{{\mathbb R}}

\def\mc{\mathcal}

\def\le{\left} \def\ri{\right}

\def\te#1{\mathrm{e}^{#1}}  \def\td{\text{\rm d}}
 
\def\I{\int}  

\def\WH{\widehat}

\def\a{\alpha} \def\b{\beta}
  \def\d{\delta}   

\def\e{\epsilon} 
\def\m{\mu} 
  \def\nn{\nonumber}   \def\s{\sigma}
\def\t{\tau}     
  \def\q{\qquad} \def\D{\Delta}
 \def\G{\Gamma}  
 \def\Th{\Theta}

\def\ie{\mathrm i}
\def\ds{\displaystyle}

\newcommand{\proof}{\no{\it Proof\ }}

\newcommand{\exit}{{\mbox{\, \vspace{3mm}}} \hfill\mbox{$\square$}}

\begin{document}

\title{A transform approach to compute\\
prices and greeks of barrier options \\ driven by
a class of L\'{e}vy processes}
\author{ Marc Jeannin$^{\dagger, \star}$ and Martijn Pistorius$^{\star}$\\
\phantom{..............}\\
$\phantom{|}^{\dagger}${\small Models and Methodology Group,
Risk Management Department}\\
{\small Nomura International plc.}\\
{\small Nomura House 1 St Martin's-le-Grand, London EC1A 4NP, U. K.}\\
\phantom{..............}\\
$\phantom{|}^{\star}${\small Department of Mathematics, Imperial College London}\\
{\small South Kensington Campus, London SW7 2AZ, U.K.}\\
{\small E-mail: m.pistorius@imperial.ac.uk}\\
\phantom{..............}
}
\date{{\small First version submitted: June 2007.\\ 
To appear in {\it Quantitative Finance}.}}

\maketitle \thispagestyle{empty}

\begin{abstract}
In this paper we propose a transform method to compute the prices
and greeks of barrier options driven by a class of L\'{e}vy
processes. We derive ana\-ly\-tical expressions for the Laplace
transforms in time of the prices and sensitivities of single
barrier options in an exponential L\'{e}vy  model with
hyper-exponential jumps. Inversion of these single Laplace
transform yields rapid, accurate results. These results are
employed to construct an approximation of the prices and
sensitivities of barrier options in exponential {\it generalised
hyper-exponential} (GHE) L\'{e}vy models. 
The latter class includes many of the
L\'{e}vy models employed in quantitative finance such as the
variance gamma (VG), KoBoL, generalised hyperbolic, and the normal
inverse Gaussian (NIG) models.  Convergence of the approximating
prices and sensitivities is proved. To provide a numerical
illustration, this transform approach is compared with Monte Carlo
simulation in the cases that the driving process is a VG and a NIG
L\'{e}vy process. Parameters are calibrated to Stoxx50E call
options.

\vfill


\textit{Acknowledgements:} This research was supported by the
Nuffield Foundation grant no. NUF-NAL/00761/G and EPSRC grant
EP/D039053/1. The research was in part carried out while the
authors were based at King's College London. We would like to
thank Kenichi Kurasawa and Denis Wallez for fruitful
conversations, Patrick Howard for his support and John Crosby,
Lane Hughston, Alex Mijatovic, and two anonymous referees for
useful suggestions that improved the paper.

\end{abstract}

\newpage{}

\setcounter{page}{1}
\section{Introduction}\label{sec:intro}
Barrier options are derivatives with a pay-off that depends on
whether a reference entity has crossed a certain boundary. Common
examples are the knock-in and knock-out call and put options that
are activated or des-activated when the underlying crosses a
specified barrier-level. Barrier and barrier-type options belong
to the most widely traded exotic options in the financial markets.
Whereas knowledge of the marginal risk-neutral distribution of the
underlying at maturity suffices to obtain arbitrage free prices of
European call and put options, the valuation of barrier options
requires specification of the risk-neutral law of the underlying
price process, as it depends on the first-passage distribution of
this process. At least as important as the calculation of prices
is the evaluation of the sensitivities of the prices to various
model-parameters (the greeks) for which the law of the process is
also required.

A class of models that has been shown to be capable of generating
a good fit of observed call and put option price data is formed by
the infinite activity L\'{e}vy models, such as extended Koponen or
KoBoL \cite{BoL}, variance gamma \cite{vgmm}, normal inverse
Gaussian \cite{BN} and generalised hyperbolic processes
\cite{EKU}. This class of models has been extensively studied and
we refer for background and further references to the books by
Boyarchenko and Levendorski\u\i\; \cite{BoyarchenkoBook}, Cont and
Tankov \cite{ContTankov} and Schoutens \cite{Schoutens}. In this
paper we consider barrier options driven by L\'{e}vy processes
with a completely monotone L\'{e}vy densities on each half-axis
(which we call {\it generalised hyper-exponential}). This class
contains many of the L\'{e}vy models used in financial modelling
as the forementioned ones and also jump-diffusions with
double-exponential jumps, as shown in Section \ref{sec:model}.

The calculation of first-passage distributions and barrier option
prices in (specific) L\'{e}vy models has been investigated in a
number of papers.  Geman and Yor \cite{GY} calculated prices and
deltas of double barrier options under the Black-Scholes model.
For spectrally one-sided L\'{e}vy processes with a Gaussian
component Rogers \cite{rogers} derived a method to evaluate
first-passage distributions. Kou and Wang \cite{KouWang}, Lipton
\cite{Lipton} and Sepp \cite{Sepp} followed a transform approach
to obtain barrier prices for a jump-diffusion with exponential
jumps. In the setting of infinite activity L\'{e}vy processes with
jumps in two directions Cont and Voltchkova \cite{ContV}
investigated discretisation of the associated integro-differential
equations; Boyarchenko and Levendorski\u\i\; \cite{BoyarchenkoL}
employed Fourier methods to investigate barrier option prices for
L\'{e}vy processes of regular exponential type; Asmussen et al.
\cite{AMP} priced an equity default swap under a CGMY model, by
fitting a hyper-exponential density to the CGMY L\'{e}vy density.
In this paper we adopt the latter approach. As first step we
obtain, in a L\'{e}vy model with hyper-exponential jumps,
analytical formulas for the Laplace transform in time of knock-in
and knock-out barrier option prices by exploiting the availability
of an explicit Wiener-Hopf factorisation. Using these results we
also establish analytical formulas for the corresponding
sensitivities with respect to the initial price (delta and gamma)
and the time of maturity (theta) up to one Laplace transform in
time. The actual barrier option prices and sensitivities are then
obtained by numerically inverting these single Laplace transforms,
using Abate and White's algorithm \cite{abate95}, yielding fast
and accurate results. Since hyper-exponential L\'{e}vy processes
are dense in the class of generalised hyper-exponential L\'{e}vy
processes, the idea is to approximate the barrier option prices
and sensitivities in a generalised hyper-exponential model by
those in an appropriately chosen hyper-exponential L\'{e}vy model.
At this point it is worth remarking that for a general L\'{e}vy
process the Wiener-Hopf factors are not available in analytically
tractable form (as they are expressed in terms of the
one-dimensional distributions that are generally not available)
and, furthermore, even if the Wiener-Hopf factors have been
obtained still a three-dimensional Laplace/Fourier inversion would
be needed to obtain the knock-in and out call option prices (see
e.g. \cite[p.372]{ContTankov}).

Following the approach described in the previous paragraph 
barrier option prices and
sensitivities under a generalised hyper-exponential L\'{e}vy model
can, at least in principle, be approximated arbitrarily closely.
Indeed, we will prove that, when a sequence of hyper-exponential
L\'{e}vy processes weakly converge to a given generalised
hyper-exponential L\'{e}vy process of infinite activity, the
corresponding first-passage times converge in distribution, and
the barrier option prices and smoothed sensitivities converge
pointwise to those of the limiting model. We illustrated this
approximation procedure by implementing it for the exponential
variance gamma (VG) and normal inverse Gaussian (NIG) models, with
parameters calibrated to Stoxx50E options. 
Using a least-squares algorithm to minimize the root mean square error 
of the approximating density with respect to the target density over a
grid, using 7 upward and downward phases, we determined the parameters of the approximating
hyper-exponential L\'{e}vy densities; the resulting
hyper-exponential L\'{e}vy processes we employed as approximations
to the VG and NIG processes. We calculated the prices and
sensitivities of European and American digital options and
down-and-out put options following this approach, and also by Monte
Carlo simulation. We found a
general agreement between the results, with relative errors in the
range of 0.5-2.5\% for prices and deltas, some distance away from
the barrier. Numerical experiments showed that closer to the
barrier errors may be larger, especially for the delta, suggesting
that a more accurate approximation of the target density by 
a hyperexponential density would be needed to reduce the 
size of the error, which could be achieved by increasing the number of phases or by employing one of the methods from the approximation 
theory of real valued functions. The phenomenon of larger 
errors in the vicinity of the barrier was also observed by Kudryatsev and Kudryatsev and Levendorski{\u\i}
\cite{KuLe} when approximating first touch digital option prices
under a NIG L\'{e}vy model using the Kou model. 
It would be
desirable to analyze the dependence of the error on the different
parameters and the distance to the barrier, and how the presented
approach compares to alternative approaches such as finite
difference schemes, but, in the interest of brevity, these
questions are left for future research.

The remainder of the paper is organized as follows. In Section
\ref{sec:model} the `generalised hyper-exponential L\'{e}vy model' is
defined and it is shown that many of the existing L\'{e}vy models 
used in quantitative finance are contained in this class. Results on the Wiener-Hopf
factorisation and first passage times for processes from this
class can be found in Section \ref{sec:WH}. Analytical results for
the Laplace transforms in time of barrier option prices are
obtained in Section \ref{sec:LTbarrier}. Using these results explicit
expressions are derived in Section \ref{sec:sens} for the Laplace
transforms of the theta, delta and gamma of the barrier an digital options.
In Section \ref{sec:numerical} numerical results are presented for
prices and sensitivities of barrier options in a VG and a NIG
model respectively, with convergence results presented in Section
\ref{sec:convsens}. Some proofs are relegated to the Appendix.

\section{Model}\label{sec:model}
Let $\{X(t), t\ge 0\}$ be a L\'{e}vy process, that is, a
stochastic process with independent and homogeneous increments,
normalised such that $X(0)=0$, defined on some appropriate
probability space $(\Omega,\mc F, P)$. 
For background on the fluctuation theory of L\'{e}vy processes and 
the application of L\'{e}vy processes in quantitative finance 
we refer to Bingham \cite{bing} and Cont and Tankov \cite{ContTankov}, 
respectively --- Sato \cite{Sato} and Bertoin \cite{Bertoin} are general treatments of the theory of 
L\'{e}vy processes. 
Assume that $X$
satisfies $E[\te{X(t)}]=\te{(r-d)t}$ for all $t\ge0$, where $r$ and $d$ are non-negative
constants representing the risk-free rate of return and the dividend 
yield, and consider
the model for the risk-neutral evolution of the stock price $S$
given by
$$
S(t) = S_0\te{X(t)}.
$$
As a consequence of the independent increments property of $X$,
$\te{-(r-d)t}S(t)$ is a martingale (under the measure $P$ and with
respect to its natural filtration).
We will restrict ourselves to the following class of L\'{e}vy
processes: \medskip

\noindent {\bf Definition.} A L\'{e}vy process is said to be {\it
generalised hyper-exponential} (and we will denote this class of processs by $GHE$) if its L\'{e}vy measure 
admits a density $k$ of the form $k(x)=k_+(x)\mbf 1_{\{x>0\}} +
k_-(-x)\mbf 1_{\{x<0\}}$ where $k_+,k_-$ are completely monotone
functions on $(0,\infty)$ and $\mbf 1_A$ denotes the indicator of
a set $A$.
\medskip

\noindent In view of Bernstein's theorem a L\'{e}vy process $X$ is a member of the class  $GHE$ if and only
if its L\'{e}vy density $k$ is of the form
\begin{equation}\label{eq:k}
k(x) = \mbf 1_{\{x>0\}}\int_0^\infty \te{-ux}\mu_+(\td u) + \mbf
1_{\{x<0\}}\int_{-\infty}^0 \te{-|ux|}\mu_-(\td u)
\end{equation}
for some measures $\mu_+,\mu_-$ on $(0,\infty)$ and $(-\infty,0)$
respectively. The mass of the interval $[a,b]$, under the measure
$\mu_+$ corresponds to the frequency of positive exponential jumps of mean
sizes between $1/a$ and $1/b$. A similar statement holds true for
the negative jumps and $\mu_-$.  In the case that the measure
$\mu_+$ is a convex combination of point-masses, the locations
$u_i>0$ and sizes $p_i>0$ of the point-masses respectively
correspond to average size $1/u_i$ and frequency $p_i$ of the
exponential jumps. Below we show that many of the L\'{e}vy
processes employed in financial modelling are generalised
hyper-exponential by explicitly calculating the corresponding
measures $\mu_\pm$. In particular, the class $GHE$ contains the time
changes of Brownian motion by a generalised hyper-exponential
subordinator.

\begin{Prop} Let $W$ be a Brownian motion and
$Y$ an independent generalised hyper-exponential subordinator. 
Then, for $\mu\in\R$, $X(t) =
W(Y(t)) + \mu Y(t)$ is a L\'{e}vy process in the class $GHE$.
\end{Prop}

\proof Denoting by $\rho$ the L\'{e}vy density of $Y$ and $\mu_Y$
the measure in the representation \eqref{eq:k},  Theorem 30.1 in
Sato (1999) implies that the L\'{e}vy density $k$ of $X$ is given
by
$$
k(x) = \int_0^\infty \frac{1}{\sqrt{2\pi s}}\te{-(x-\mu
s)^2/(2s)}\rho(s)\td s.
$$
An interchange of the order of integration yields that
\begin{eqnarray*}
k(x) &=& \int_0^\infty \frac{1}{\sqrt{2\pi s}}\te{-(x-\mu
s)^2/(2s)}\int_0^\infty
\te{-su}\mu_Y(\td u)\td s\\
&=& \int_0^\infty \int_0^\infty \frac{1}{\sqrt{2\pi s}}\te{-(x-\mu
s)^2/(2s)}\te{-s u}\td s
\mu_Y(\td u)\\
&=& \int_0^\infty \frac{1}{\sqrt{\mu^2 + 2u}}\te{-|x|\sqrt{\mu^2 + 2u}
+ \mu x}\mu_Y(\td u),
\end{eqnarray*}
and the claim follows.\exit \medskip

\noindent The following result gives necessary and sufficient
conditions on the measures $\mu_+, \mu_-$ such that $k$ in
\eqref{eq:k} is a L\'{e}vy density:

\begin{Prop} Let $\mu$ be a measure on $\R\backslash\{0\}$ and set
$\mu_\pm(\td x) = \mbf 1_{\{\pm x>0\}}\mu(\td x)$. Then $k$
defined in \eqref{eq:k} is a L\'{e}vy density if and only if
\begin{equation}\label{eq:m+int}
\int \frac{1}{|u|}\wedge\frac{1}{|u|^3}\; \mu(\td u) < \infty.
\end{equation}
\end{Prop}

\proof By interchanging the order of integration it can be
verified that
\begin{eqnarray*}
\int_1^\infty\int_0^\infty \te{-ux}\mu(\td u)\td x &=& \int_0^\infty
\frac{\te{-u}}{u}\mu(\td u),\\
\int_0^1x^2\int_1^\infty \te{-ux}\mu(\td u)\td x &=& \int_1^\infty
\frac{2}{u^3} -  \frac{\te{-u}}{u}\le(1 + \frac{2}{u} +
\frac{2}{u^2} \ri)\mu(\td u).
\end{eqnarray*}
Further we note that $\int_0^1x^2\int_0^1 \te{-ux}\mu(\td u)\td x$
is bounded below and above by $\frac{\mu(0,1)}{3\mrm e}$ and
$\frac{\mu(0,1)}{3}$, respectively. In view of these calculations
we read off that $k$ in \eqref{eq:k} satisfies the integrability
condition $\int [1\wedge x^2] k(x)\td x < \infty$ (which is the
requirement for $k$ to be a L\'{e}vy density) if and only if
\eqref{eq:m+int} holds.\exit
\smallskip

\no In the following examples we explicitly determine the measures
$\mu_\pm$.
\medskip

\noindent{\bf Examples.} We shall write $k_+$ for the density $k$
restricted to the positive half-axis.

\medskip

\no\ $\bullet$
{\bf Double exponential model} (Kou \cite{Kou})
For a jump-diffusion model where the jumps follow a double exponential
distribution, $\mu_+$ is a point-mass located at the reciprocal of the
mean jump sizes:
$$k_{+}(x) = \lambda^+\a^+\te{-\a^+ x},\q\q
\mu_+(\td u)=\lambda^+\a^+\delta_{\a^+}(\td u),$$
where $\a^+,\lambda^+>0$.

\no\ $\bullet$ {\bf Hyper-exponential model} (e.g. \cite{AAP})
This model is an extension
of Kou's model by allowing for exponential jumps with
a finite number of different means. For
$p^+_i,\a^+_i,$ and $\lambda^+>0$ with $\sum_{i=1}^np^+_i=1$ we thus get
$$
k_{+}(x) = \lambda^+\sum_{i=1}^np^+_i\a_i^+\te{-\a^+_i x},\q\q
\mu_+(\td u)=\lambda_+\sum_{i=1}^np^+_i\a_i^+\delta_{\a^+_i}(\td u).
$$

\no\ $\bullet$ {\bf KoBoL/CGMY model} (\cite{BoL, CGMY}) In the
KoBoL model (also called CGMY model), the measure $\mu$ has a
continuous density $k$, and it holds that
$$k_{+}(x) = \frac{C}{|x|^{Y+1}}\te{-M |x|},
\q\q
\mu_+(\td u)= C \frac{(u-M)^{Y}}{\G(1+Y)}\mbf 1_{\{u\ge M\}}\td u,
$$
where $C, M, Y>0$. The form of $\mu_+$ invokes the definition of
the gamma function
$$
\frac{1}{x^{Y+1}} = \int_0^\infty \te{-ux} \frac{u^Y}{\G(1+Y)}\td u.
$$
In particular, for the variance gamma model, we set $Y=0$ and get
$k_{+}(x) = Cx^{-1}\te{-M x}$ and $\mu_+(\td u)= C\mbf 1_{\{u\ge
M\}}\td u$.
\medskip

\no\ $\bullet$ {\bf Meixner model} (e.g. \cite{Schoutens}) The
L\'{e}vy density of a Meixner L\'{e}vy process is given by
$$
k_{+}(x) =\frac{\d\te{\b x/\a}}{x\sinh(\pi x/\a)} =
2\te{\beta x\/a - \pi|x|/\a}\sum_{n=0}^\infty
\frac{\te{-2\pi n|x|/\a}}{|x|}, \quad x\neq 0,
$$
where $\d,\a>0$, $-\pi<\b<\pi$, so that we find that
$$
\mu_+(\td u) = 2\d\sum_{n=0}^\infty
\mbf 1_{\{u\ge (2\pi n + \pi - \beta)/\a\}}\td u.
$$

\no\ $\bullet$ {\bf Normal-inverse Gaussian (NIG)} (Barndorff
Nielsen \cite{BN}) In the NIG model, the measure $\mu$ has a
density which reads as
$$k_+(x) = \frac{C\d\a}{\pi}\te{\b x}\frac{K_1(\a x)}{x},
\q \mu_+(\td u) = \frac{C\d\a}{\pi}([u+\b]/\a)^2-1)^{1/2}\mbf
1_{\{u\ge \a-\b\}}\td u,
$$
where $\a>\b>0$, $\d, C > 0$ and $K_1$ is the McDonald function. The form of $\mu_+$ is based on
the following representation (see \cite{AbrStegun}) of $K_1$
$$
K_1(x) = x \int_1^\infty \te{-xv}(v^2 - 1)^{1/2}\td v.
$$

\no\ $\bullet$ {\bf Generalised hyperbolic (GH)} (Eberlein et al.
\cite{EKU}) The GH process can be respresented as time change of
Brownian motion by a generalised inverse Gaussian (GIG)
subordinator. The L\'{e}vy density of GIG is a generalised gamma
convolution which means in particular that it is of the form
\eqref{eq:k}.

\bigskip

\section{First passage theory}\label{sec:WH}
First passage distributions are an essential element in the
valuation of barrier options. In this section we briefly review
elements of the first passage theory for L\'{e}vy processes that
will be needed in the sequel.

\subsection{Wiener-Hopf factorisation}
The distributions of the running supremum and the running infimum
of $X$ are linked to the distribution of $X(t)$ via the famed
Wiener-Hopf factorisation of the L\'{e}vy exponent of $X$, denote
by $\Psi(u) = \log E[\te{\ie u X(1)}]$. For $v\in\mathbb C$, 
let $\Im(v)$ denote the imaginary part of $u$.
\medskip

\noindent{\bf Definition.} A {\it Wiener-Hopf
factorisation of} $\Psi$ is a pair of functions $F^+_q, F^-_q$ that
satisfies, for $u\in\mathbb R$ and $q>0$, the relation
\begin{equation}
q(q - \Psi(u))^{-1} = F^+_q(u)F^-_q(u),
\end{equation}
where, for fixed $q>0$, $u\mapsto F^\pm_q(u)$  
are continuous and non-vanishing
on $\pm \Im(u)\ge 0$ and analytic
on $\Im(u)>0$ with $F_q^\pm(0)=1$, and
\begin{equation}\label{eq:qi}
\lim_{q\to\infty} F_q^\pm(u) = 1.
\end{equation}

Denote by $\ovl X(t)$ and $\unl X(t)$ the running supremum and
infimum of $X$ up to time $t$,
$$
\ovl X(t) = \sup_{s\leq t} X(s),\quad\quad \unl X(t) = \inf_{s\leq t}
X(s),
$$
and let $q^{-1}\Phi^+_q$, $q^{-1}\Phi^-_q$ be the respective
Fourier-Laplace transforms
\begin{equation*}
\Phi^+_q(u) = \int_0^\infty q\te{-q t}E[\te{\ie u \ovl X(t)}]\td
t,\quad\quad \Phi^-_q(u) = \int_0^\infty q\te{-q t}E[\te{\ie u \unl
X(t)}]\td t.
\end{equation*}
It is not hard to verify that $\Phi_q^\pm$ satisfy the condition \eqref{eq:qi}.
Rogozin \cite{rogozin66} established the following result:

\begin{Thm}[Rogozin]
$\Phi^+_q(u), \Phi^-_q(u)$ is the unique Wiener-Hopf factorisation of $\Psi$.
\end{Thm}
If the L\'{e}vy measure has support in both half-axis the
factorisation is generally not explicitly known, except if the
L\'{e}vy measure of $X$ is of a particular form.
The observation that the Wiener-Hopf factorisation and related
first passage distributions are tractable for a random walk in the
case that the jump distribution on the positive half-axis follows
an exponential distribution can already be found in Feller
\cite{fel71}. See also the books Borovkov \cite{BorovkovQ} and
Asmussen \cite{AsmussenQ} for background on the Wiener-Hopf
factorisation for random walks. Mordecki \cite{mord02} and Lipton
\cite{Lipton} calculated the Laplace transforms of first passage
time distributions for a L\'{e}vy process with hyper-exponential
jumps. Asmussen et al. \cite{AAP} derived explicitly the
Wiener-Hopf factorisation if the jump distributions of the
L\'{e}vy process are of phase-type.

In the sequel we will employ the known factorisation results for a
jump-diffusion with L\'{e}vy density (also to be called a
{\it hyper-exponential jump-diffusion (HEJD)}) given by
\begin{equation}\label{eq:ka}
k(x) = \lambda^+\sum_{i=1}^{n^+}p_i^+\a_i^+\te{-\a_i^+x}\mbf
1_{\{x>0\}} + \lambda^-\sum_{j=1}^{n^-}p_j^-\a_j^-\te{-\a_i^-x}\mbf
1_{\{x<0\}},
\end{equation}
where $\a_i^\pm, \lambda^{\pm} p_i^+>0$ with $\sum_{i=1}^{n^\pm}p_i^\pm=1$.
The corresponding L\'{e}vy exponent then reads as
\begin{equation}\label{eq:PSI}
\Psi(u) = \m u\ie - \frac{\s^2}{2}u^2 +
\lambda^+\sum_{i}p_i^+\le(\frac{\a_i^+}{\a_i^+ - u\ie} - 1\ri) +
\lambda^-\sum_{j}p_j^-\le(\frac{\a_j^-}{\a_j^- + u\ie} - 1\ri).
\end{equation}
The function $\Psi$ in \eqref{eq:PSI} 
can be analytically extended to the complex plane
except the finite set $\{-\ie\a^+_i, \ie\a^-_i, i=1, \ldots, n^\pm\}$, and this extension will also be denoted by $\Psi$. 
Denoting by $\rho^+_i = \rho^+_i(q)$,
$i=1,\ldots, m^+$ and $\rho^-_j = \rho^-_j(q)$, $j=1,\ldots, m^-$ the
roots  of $\Psi(- \ie s)=q$ with positive and negative real parts respectively,
the Wiener-Hopf factors $\Phi_q^+$ and $\Phi_q^-$ are explicitly given as follows:
\begin{eqnarray*}
\Phi_q^+(u) = \frac{\prod_{i=1}^{n^+}\left(1 -
\frac{u\ie}{\alpha^+_i}\right)}{\prod_{i=1}^{m^+}
\left(1 - \frac{u\ie}{\rho^+_i(q)}\right)},\qquad
\Phi_q^-(u) = \frac{\prod_{j=1}^{n^-}\left(1 +
\frac{u\ie}{\alpha^-_j}\right)}{\prod_{j=1}^{m^-}
\left(1 - \frac{u\ie}{\rho^-_j(q)}\right)}.
\end{eqnarray*}
Note that as a consequence of the intermediate value theorem and the fact that $s\mapsto\Psi(-\ie s)$ is a rational function with real coefficients and single poles, it follows that the roots of the equation 
$\Psi(- \ie s)=q$ are all real and distinct if $q\in\R_+$.
Performing a partial fraction decomposition and termwise inverting
the terms yields that the time-Laplace transforms of the distributions
of $\ovl X(t)$ and $\unl X(t)$ are given by\footnote{In the case that there are multiple roots, which might be the case for some complex values of 
$q$, it is still possible to perform a partial fraction decomposition 
which results in similar but slightly different expressions -- 
see Remark 4 in \cite{AAP}. }

\begin{equation}  \left \{ \begin{array}{rl}\centering
\ds \int_0^\infty \te{-qt} P(\ovl X(t) \leq z)\td t& =\ds
\frac{1}{q}\le(1 - \ds\sum_{i=1}^{m^{+}}A_{i}^{+}\te{-\rho^+_i(q)
z}\ri),
\quad\quad  z\ge 0,\\
\\
\ds \int_0^\infty \te{-qt}P(-\unl X(t)\leq z)\td t &=\ds
\frac{1}{q}\le( 1 - \ds \sum_{j=1}^{m^{-}}A_{j}^{-}\te{\rho^-_j(q)
z}\ri), \quad\quad z\ge 0,
\end{array}\ri.
\end{equation}
where the coefficients $A_i^+$ and $A_j^-$ are given by
$$
A_i^+ = \frac{\prod_{v=1}^{n^+}\left(1 -
\frac{\rho^+_i(q)}{\alpha^+_v}\right)}{\prod_{v=1,v\neq i}^{m^+}\left(1 -
\frac{\rho^+_i(q)}{\rho^+_v(q)}\right)},\quad%
A_j^- = \frac{\prod_{v=1}^{n^-}\left(1 +
\frac{\rho^-_j(q)}{\alpha^-_v}\right)}{\prod_{v=1,v\neq j}^{m^-}\left(1 -
\frac{\rho^-_j(q)}{\rho^-_v(q)}\right)}.
$$

\subsection{Generalised hyper-exponential L\'{e}vy processes}\label{sec:genhyp}
For a generalised hyper-exponential L\'{e}vy process $X$ the
Wiener-Hopf factorisation is in general not known explicitly.
However, due to its special form \eqref{eq:k}, its L\'{e}vy
density can be approximated arbitrarily closely
by a hyper-exponential density (by approximating the
measures $\mu_\pm$ by sums of point-masses), and for the resulting
hyper-exponential L\'{e}vy process the Wiener-Hopf factorisation 
is explicitly given in the previous section. As hyper-exponential 
L\'{e}vy processes are `dense' in the class $GHE$, 
the Wiener-Hopf factorisation of $X$ can be approximated arbitrarily 
accurately using this idea.

More formally, a sequence of approximating processes $(X^{(n)})_n$ can be explicitly constructed as follows. Denote by $k$ and $\s^2$ the L\'{e}vy
density and the diffusion coefficient of $X$, respectively. 
Fix a sequence $(\varepsilon_n)_n$ of positive numbers converging to zero,
and, 
for every $n$, let $(u_i)_i=(u^{(n)}_i)_i$ and 
$(v_j)_j=(v_j^{(n)})_j$  be two finite partitions of $(0,\infty)$
with vanishing mesh\footnote{A partition $(t_i)_{i=1}^m$ of $(0,\infty)$ is an
increasing set of times $0<t_1<t_2 \ldots$. The mesh of $(t_i)$ is
defined as $\max_{1\leq i\leq m}|t^{}_i - t^{}_{i-1}|$.} and 
let $(\Delta^+_i)_i = (\Delta^{+(n)}_i)_i$, 
$(\Delta^-_j)_j = (\Delta^{-(n)}_j)_j$ be two finite sets of positive 
weights, shortly to be determined. Set the approximating density equal to
\begin{equation}\label{eq:kn}
k_n(y) = \mbf 1_{\{y>0\}}\sum_i \te{-y u_i}\Delta^+_i  + \mbf
1_{\{y<0\}} \sum_j \te{y v_j}\Delta^-_j.
\end{equation}
For a given $n$, the partitions $(u_i)_i$ and $(v_j)_j$ and the weights $(\Delta_i^+)_i$, $(\Delta_j^-)_j$ are chosen such that the mass of $k$ in the tails and the $L^2$ distance between the target density $k$ and the approximating density $k_n$ on a closed bounded set not containing zero is smaller than $\varepsilon_n$, that is,
$$\int_{\R\backslash A_n} k(y)\td y< \varepsilon_n,
\qquad 
\int_{A_n\backslash B_n}(k(y) - k_n(y))^2\td y < \varepsilon_n,
$$
for some open bounded set $A_n, B_n$ with 
$0\in B_n\subset A_n\subset \mathbb R$.
These requirements can be fulfilled since
$k$ has finite mass on the set $\{|y|>1\}$ and takes the
special form \eqref{eq:k}. Set $X^{(n)}$ equal to the L\'{e}vy process with
L\'{e}vy density $k_n$ and Gaussian coefficient $\s^2_n$
\begin{equation}\label{eq:sn}
\s^{2}_n = \s^2 + \int_{-v^{(n)}_1}^{u^{(n)}_1}y^2(k(y) -
k_n(y))_+\td y
\end{equation}
and with drift $\m_n$ chosen such that $E[\te{X^{(n)}(1)}] = \te{r-d}$.
The approximating processes $X^{(n)}$ constructed in this way
can be shown to converge weakly to $X$.
\footnote{This convergence takes
place in the space $D(\R_+)$
endowed with the Skorokhod topology. A proof of this fact is given in Appendix \ref{sec:weak}, Lemma \ref{lem:weak}.}
Convergence of the barrier option prices corresponding to log price $X^{(n)}$
rests on the fact\footnote{A proof is given in Appendix \ref{sec:weak}, Lemma \ref{lem:weakXX}.} that the weak
convergence of $X^{(n)}$ to $ X$ carries over to convergence of the
running supremum, infimum and the
crossing times
$$
T^{(n)}(x) = \inf\{t\ge 0: X^{(n)}(t)\leq x\}\q\text{ and }\q T(x)
= \inf\{t\ge 0: X(t)\leq x\},
$$
as summarised in the following result.

\begin{Prop}\label{prop:convWH}
Let $X$ be generalised hyper-exponential and let $X^{(n)}$ be as
defined. For $T>0$, $(X^{(n)}(T), \ovl {X}^{(n)}(T))$ converges in
distribution to $(X(T), \ovl X(T))$ as $n\to\infty$. In
particular, if $X$ is not of type A,\footnote{A L\'{e}vy process
is called of type A (see Sato(1999),p. 65) if its L\'{e}vy measure
has finite mass and no Brownian motion is present. Type A L\'{e}vy
processes are the compound Poisson processes added to a (possibly
zero) deterministic drift.} it holds for $t>0$ and $z>0$ that
\begin{eqnarray}
\label{conv1} P(\ovl X^{(n)}(t) \leq z) &\to& P(\ovl X(t)\leq
z),\quad\quad
P(-\unl X^{(n)}(t)\leq z) \to P(-\unl X(t) \leq z),\\
P(T^{(n)}(-z) \leq t) &\to& P(T(-z) \leq t),
\label{conv2}
\end{eqnarray}
as $n\to\infty$.
\end{Prop}

\section{Laplace transforms of barrier option  prices}\label{sec:LTbarrier}
In this section we employ the Wiener-Hopf factorisation results to derive
the values of single barrier options in a hyper-exponential L\'{e}vy model.
We will restrict ourselves to  down-and-out and
down-and-in digital and put options, but related options
(such as up-and-in call options) can be derived similarly.
\subsection{Digital options}
A digital option is a derivative that pays $1$ euro if the price
of the underlying asset has up or down-crossed a level $H$ before
a maturity time $T$, with pay-off occuring either directly at the
moment of crossing or at the maturity time $T$. We will derive the
values of down-and-out and down-and-in digital options and denote
by $EDOD$ and $EDID$ the respective prices corresponding to
payment at maturity. We also consider the American version of the
down-and-out digital option (denoted by $ADID$) that pays out at
the moment of crossing the barrier. The following result gives the
Laplace transforms of the digital prices when $S_0>H$. Denote by
$\WH f$ the Laplace transform of $f$:
$$
\WH f(q) = \int_0^\infty \te{-qt}f(t)\td t.
$$
\begin{Prop}\label{prop:digital}
If $q>0$ and $H < S_0$ it holds that
\begin{eqnarray*}
\WH{EDID}(q) &=& \frac{1}{q+r}\sum_{j=1}^{k^-}
A_j^-\le(\frac{S_0}{H}\ri)^{\rho^-_j},\\
\WH{ADID}(q) &=& \frac{q+r}{q} \WH{EDID}(q),\quad\quad
\WH{EDOD}(q) = \frac{1}{q+r} - \WH{EDID}(q).
\end{eqnarray*}
where $\rho_j^- = \rho^-_j(q+r)$.
\end{Prop}
{\bf Proof of Proposition \ref{prop:digital}:} Since the
riskless rate of return is assumed to be constant and equal to $r$, it follows
by standard no-arbitrage pricing arguments that the price of the
down-and-in digital option is given by
\begin{equation}\label{eq:EDID}
EDID(T) = EDID(T,S_0) = \te{-rT}P[\unl X(T)\leq \log(H/S_0)],
\end{equation}
where we switched to log-scale. Also, if we denote $h=\log(H/S_0)$
then
$$
ADID(T) = ADID(T,S_0) = E[\te{-rT(h)}\mbf 1_{\{\unl X(T)\leq
h\}}].
$$
Taking now the Laplace transforms of these expressions in $T$ and
combining with the first-passage distributions given in Section
\ref{sec:WH} we find if $h<0$:
\begin{eqnarray*}
\WH{EDID}(q) &=& \I_0^\infty\te{-(q+r)T}P[\unl X(T) \leq h]\td T
= \I_0^\infty\te{-(q+r)T}P[T(h) \leq T]\td T \\
&=& \frac{1}{q+r}\int_{-h}^\infty\sum_{j=1}^{k^-}A_j^-(-\rho_j^-)\te{\rho_j^-
y}\td y = \frac{1}{q+r}\sum_{j=1}^{k^-}A_j^-\te{-\rho_j^- h},\\
\WH{ADID}(q) &=& \I_0^\infty\te{-qT} E[\te{-rT(h)}\mbf 1_{\{\unl
X(T) \leq H\}}]\td T
 = E\le[\I_0^\infty\te{-qT-rT(h)}\mbf 1_{\{T(h)\leq T\}}\td t\ri]\\
&=& 
\frac{1}{q}E[\te{-(q+r)T(h)}] =
\frac{q+r}{q}\I_0^\infty\te{-qT}\{\te{-rT}P[T(h) \leq T]\}\td T.
\end{eqnarray*}
The last identity follows by noting that $EDOD = \te{-rT} - EDID$.
\exit


\subsection{Knock-out and knock-in put options}
A down-and-out put pays out at maturity $T$ the strike $K$ less
the value of the asset $S_T$ (if this difference is positive) with
the added feature that the put is worthless if the price of the
asset has been below a level $H$ by time $T$. In the next result
we derive the Laplace transform in the maturity $T$ of
down-and-out put $DOP(T)=DOP(T;S_0,K,H)$. The value of the down-and-in
put is presented in Appendix \ref{app:DIP}.

\begin{Prop}\label{prop:barrier} Let $q>0$ and write
$h=\log (H/S_0), \ell=\log(K/S_0)$. Then it holds that
$$
\WH{DOP}(q) = \frac{1}{q+r}KC^{(0)}(\ell,h) -
\frac{S_0}{q+r}C^{(1)}(\ell,h),
$$
where, if $h<\ell<0$,
\begin{eqnarray*}
C^{(b)}(\ell,h) &=& \sum_{j=1}^{k^-}\sum_{i=1}^{k^+}
\frac{\rho_i^+(-\rho_j^-)}{(\rho_j^--\rho_i^+)(b-\rho_i^+)}A_i^+A_j^-
(\te{(b-\rho_i^+)\ell+(\rho_i^+-\rho_j^-)h}-\te{(b-\rho_j^-)\ell})\\
&+& \le(1 - \sum_{i=1}^{k^+}\frac{b}{b-\rho_i^+}A_i^+\ri)
\sum_{j=1}^{k^-}A_j^-\frac{(-\rho_j^-)}{\rho_j^--b}(\te{(b-\rho_j^-)h} -
\te{(b-\rho_j^-)\ell})
\end{eqnarray*}
and, if $h<0<\ell$,
\begin{eqnarray*}
C^{(b)}(\ell,h) &=& \sum_{j=1}^{k^-}\sum_{i=1}^{k^+}
\frac{\rho_i^+(-\rho_j^-)}{(\rho_j^--\rho_i^+)(b-\rho_i^+)}A_i^+A_j^-
\te{(b-\rho_i^+)\ell}(\te{(\rho_i^+-\rho_j^-)h}- 1)\\
&+&
\le(1 - \sum_{i=1}^{k^+}\frac{b}{b-\rho_i^+}A_i^+\ri)\le(1
+ \sum_{j=1}^{k^-}A_j^-\le(\frac{(-\rho_j^-)}{\rho_j^--b}\te{(b-\rho_j^-)h} -
\frac{b}{\rho_j^--b}\ri)\ri)\\
&+& \le(1 - \sum_{j=1}^{k^-}A_j^-\ri)
\sum_{i=1}^{k^+}\frac{\rho_i^+}{b-\rho_i^+}A_i^+\te{(b-\rho_i^+)\ell},
\end{eqnarray*}
with $\rho^+_i=\rho^+_i(q+r)$ and $\rho^-_j=\rho^-_j(q+r)$.
\end{Prop}
It is easy to check that, in the case of exponential upward and
downward jumps ($n^\pm = 1$), this result agrees those obtained
before by Kou and Wang \cite{KouWang} and Sepp \cite{Sepp}.

\no{\bf Proof of Proposition \ref{prop:barrier}:} By the standard
theory of no-arbitrage-pricing it follows that an arbitrage free
price for the down-and-out barrier option is given by
\begin{eqnarray}\nn
DOP(T;S_0,K,H) &=& \te{-rT}E[(K-S(T))^+\mbf 1_{\{\unl S(T) > H\}}]\\
\nn &=& \te{-rT}KP[\unl X(T) > h, X(T)<\ell] \\ &\phantom{=}& -
\te{-rT}S_0E[\te{X(T)}\mbf 1_{\{\unl X(T) > h, X(T)<\ell\}}],
\label{eq:DOP}
\end{eqnarray}
where $h=\log(H/S_0)$ and $\ell=\log(K/S_0)$. Denoting by $\tau =
\tau(q+r)$ an independent random time with an exponential
distribution with parameter $q+r$, the Laplace transform
$\WH{DOP}(q)$ of $DOP(T;S_0,K,H)$ can be compactly represented as
\begin{eqnarray}
\nn\WH{DOP}(q) &=& K\int_0^\infty\te{-(q+r)T}P[\unl X(T) > h, X(T)<\ell]\td T\\
\nn
&\phantom{=}& -  S_0\int_0^\infty\te{-(q+r)T}E[\te{X(T)}\mbf 1_{\{\unl X(T) > h,\ X(T)<\ell\}}]\td T\\
&=& \frac{K}{q+r} P[\unl X(\t) > h,\ X(\t)<\ell] -
\frac{S_0}{q+r}E[\te{X(\t)}\mbf 1_{\{\unl X(\t) > h,\
X(\t)<\ell\}}]. \label{eq:WHDOP}
\end{eqnarray}
To calculate the quantities
\begin{eqnarray*}
C^{(b)}(h,\ell) &:=& E[\te{bX(\t)}\mbf 1_{\{\unl X(\t) > h,\
X(\t)<\ell\}}]\\
g(y,b) &:=& E[\te{b\ovl X(\t)}\mbf 1_{\{\ovl X(\t) < y\}}]
\end{eqnarray*}
we shall appeal to the following two key properties: 
\BEN
\im[(a)] $\unl
X(\t)$ and $X(\tau) - \unl X(\t)$ are independent and 
\im[(b)] the pairs $(\ovl
X({\t}), \ovl X({\t}) - X({\t}))$ and $(X({\t}) - \unl X({\t}),-\unl
X({\t}))$ are identically distributed. 
\EEN
In view of (a) it follows
that
\begin{eqnarray*}
C^{(b)}(h,\ell) 
&=& E[\te{b(X(\t)-\unl X(\t))}\te{b\unl X(\t)}
\mbf 1_{\{-\min\{k,0\} < -\unl X(\t) < -h\}}\mbf 1_{\{X(\t)-\unl X(\t) + \unl X(\t)<k\}}]\\
&=& \int_{-\min\{k,0\}}^{-h}\te{-by} E[\te{b(X(\t)-\unl X(\t))}\mbf
1_{\{X(\t)-\unl X(\t)<k+y\}}] P(-\unl X(\t)\in\td y)\\
&=& \int_{-\min\{\ell,0\}}^{-h} \te{-by} g(\ell+y,b) f_{-\unl
X(\t)}(y)\td y + g(\ell,b)P[\unl X(\t) = 0],
\end{eqnarray*}
where we used property (b) in the last line. Inserting the
explicit form of the distribution of $\ovl X_\t$, we find that for
$y>0$
\begin{eqnarray*}
g(y,b) &=& \int_{0+}^{y}\te{bz}f_{\ovl X(\t)}(z)\td z
+ P[\ovl X(\t) = 0]\\
&=& \sum_{i=1}^{k^+}\frac{\rho_i^+}{b-\rho_i^+}A_i^+(\te{(b-\rho_i^+)y} - 1) +
1 - \sum_{i=1}^{k^+}A_i^+.
\end{eqnarray*}
The next step is to integrate $g(\ell+y,b)$ against $f_{-\unl
X(\t)}$. If $\ell<0$, this yields
\begin{eqnarray*}
C^{(b)}(h,\ell) &=& \sum_{j=1}^{k^-}\sum_{i=1}^{k^+}
\frac{\rho_i^+(-\rho_j^-)}{b-\rho_i^+}A_i^+A_j^-\int_{-\ell}^{-h}(\te{(b-\rho_i^+)(\ell+y)+(\rho_j^--b)y}
- \te{(\rho_j^- -b)y})\td y\\
&+& \le(1 - \sum_{i=1}^{k^+}A_i^+\ri)\sum_{j=1}^{k^-}A_j^-(-\rho_j^-)
\int_{-\ell}^{-h}\te{(\rho_j^--b) y}\td y,
\end{eqnarray*}
and in the case that $\ell>0$, it holds that
\begin{eqnarray*}
C^{(b)}(h,\ell) &=& \sum_{j=1}^{k^-}\sum_{i=1}^{k^+}
\frac{\rho_i^+(-\rho_j^-)}{b-\rho_i^+}A_i^+A_j^-\int_{0}^{-h}(\te{(b-\rho_i^+)(\ell+y)
+ (\rho_j^--b)y}
- \te{(\rho_j^--b) y})\td y\\
&+& \le(1 - \sum_{i=1}^{k^+}A_i^+\ri)\sum_{j=1}^{k^-}A_j^-(-\rho_j^-)
\int_{0}^{-h}\te{(\rho_j^--b) y}\td y \\
&+& \le(1 - \sum_{j=1}^{k^-}A_j^-\ri)
\le(\sum_{i=1}^{k^+}\frac{\rho_i^+}{b-\rho_i^+}A_i^+(\te{(b-\rho_i^+)\ell} -
1) + 1 - \sum_{i=1}^{k^+}A_i^+\ri).
\end{eqnarray*}
It is then a straightforward matter of some calculus to arrive at
the stated expressions.\exit

\section{Calculating sensitivities}\label{sec:sens}
The analytical formulas for the barrier and digital prices in the
previous section suggest the possibility of calculating the
corresponding sensitivities by direct differentiation. In this
section we make this idea precise by providing regularity results.
The delta and gamma sensitivities of the option price with respect
to the initial stock price $S_0$ will be denoted by
$$
\D_V = \frac{\partial V}{\partial S_0},\quad\quad
\Gamma_V = \frac{\partial^2 V}{\partial S_0^2},
$$
where $V=V(T,S_0)$ is the value function of the option under
consideration (digital or down-and-out put). The sensitivity
$\Theta_V$ with respect to the maturity $T$ (the theta) is defined
by a function that satisfies $$V(T) - V(0) =
\int_0^T\Theta_V(s)\td s,$$ for all $T>0$. If $V$ is continuously
differentiable with respect to $T$ then this definition is
equivalent to $\Theta_V = \frac{\partial V}{\partial T}.$ The
first result in this direction concerns the theta of a digital
option:
\begin{Lemma}\label{lem:sens}
Let $X$ be a general L\'{e}vy process not of type A. Then, for
$q>0$, $S_0 > H$ and $T>0$,
\begin{eqnarray*}
\WH\Theta_{EDID}(q) &=& q \WH{EDID}(q),\\
\Theta_{ADID}(T) &=& \Theta_{EDID}(T) +
r EDID(T).
\end{eqnarray*}
\end{Lemma}
\proof Since the map $F:[0,\infty)\to[0,\infty)$ given by $F(t) =
P(T(x)\leq t)$ is an increasing function with $F(0) = 0$ it holds
that
\begin{equation}\label{eq:LSTF}
\int_0^\infty \te{-q t} \td F(t) = q \int_0^\infty \te{-q t} F(t)\td t.
\end{equation}
Further, Lemma 49.3 in Sato(1999) implies that $P(T(h) = t) = 0$
for $h<0$, $t>0$, for L\'{e}vy processes $X$ not of type A, so
that $F$ is continuous. If we denote by $f$ a density of $F$ it
follows that $\WH f(q) = q\WH F(q)$. In view of the form of the
pay-off of the digital we deduce that $\WH\Theta_{EDID}(q) =
(q+r)\WH{EDID}(q) - r\WH{EDID}$ and the assertion follows. The
second equation follows from the relation between $EDID$ and
$ADID$ given in Proposition \ref{prop:digital}.\exit

\medskip
In the case of a Gaussian component $\sigma>0$ general results are
available in the literature regarding the smoothness and
regularity of the solutions of partial integro-differential
equations that imply, by the Feynman-Kac representation, that the
value functions of the barrier options are smooth (see Bensoussan
\& Lions \cite{Bensoussan} and Garroni \& Menaldi \cite{Garroni}).
More specifically, for a hyper-exponential L\'{e}vy process with
$\sigma>0$ it holds that $F$ with $F:(t,h)\mapsto P(\unl X(t) \leq
h)$ is element of $C^{1,2}((0,\infty)\times (-\infty,0))$ and
$G:(t,h,k)\mapsto P(\unl X(t) > h, X(t) < k)$ satisfies $G\in
C^{1,2,2}((0,\infty)\times(-\infty,0)^2)$. In view of the
probabilistic representations \eqref{eq:EDID} and \eqref{eq:DOP},
this result directly implies that the delta and gamma of $EDID$
and $DOP$ are well defined. The Laplace transforms of the delta
and gamma of $EDID$ are given as follows:

\begin{Prop}\label{prop:DDstar}
Suppose $X$ is a hyper-exponential L\'{e}vy process
with $\s>0$ and let $q>0$ and $H<S_0$. It holds that
\begin{eqnarray*}
\WH\Delta_{EDID}(q) &=& \frac{1}{q+r}\frac{1}{S_0}
\sum_{j=1}^{k^-}\rho_j^-A_j^-(S_0/H)^{\rho_j^-},\\
\WH\Gamma_{EDID}(q) &=& \frac{1}{q+r}\frac{1}{S_0^2}
\sum_{j=1}^{k^-}\rho_j^-(\rho_j^--1)A_j^-(S_0/H)^{\rho_j^-}.
\end{eqnarray*}
\end{Prop}
\proof Letting $H<c<S_0$ the fundamental theorem of calculus implies that
\begin{eqnarray*}
\frac{\partial}{\partial S_0} \WH{EDID}(q) &=& \frac{\partial}{\partial S_0}
\int_0^\infty \int_c^{S_0}\D_{EDID}(t,y)\td y\td t\\
&=& \frac{\partial}{\partial S_0}
\int_c^{S_0}\int_0^\infty \D_{EDID}(t,y)\td t \td y  = \WH{\Delta}_{EDID}(q),
\end{eqnarray*}
where the change of the order of integration is justified
(by Fubini's theorem) since $\D_{EDID}\ge 0$
and the last equality follows since $\WH{EDID}(q)=\WH{EDID}(q;S_0)$ is
continuously differentiable as a function of $S_0$.
The form of the delta follows now
from Proposition \ref{prop:digital}. The result for the gamma follows
by a similar reasoning as above, where the interchange of integration is
now justified since $\te{-qt}|\Gamma_{EDID}(t,S_0)|$ is
dominated by an integrable function (on $(0,\infty)\times(c,S_0)$).

To show the latter fact we note first that since $EDID$ is smooth enough
it satisfies the partial integro-differential equation
that reads in terms of the Greeks of $EDID$ as
$$
\le[\Theta + \mbox{$\frac{\s^2}{2}$}S^2\Gamma + \mu S\Delta -
(\lambda_+ + \lambda_-+r)EDID\ri](t,S) +
I(t,S) = 0,
$$
for $t>0, S>H>0$ with boundary values $EDID(t,S) = 1$ for $S\leq H, t\in(0,T]$
and $EDID(0,S) = 1$ if $S\leq H$ and zero else. Here $I$ is the non-local
part of the generator, given by
$$I(t,S) = \int_{-\infty}^\infty EDID(t,S\te{y})k(y)\td y.$$
Since $EDID\leq 1$ and the measure $k(y)\td y$ has finite mass,
it thus follows that
\begin{equation*}
|\Gamma_{EDID}(t,S)| \leq C(1 + |\Delta_{EDID}(t,S) | +
|\Theta_{EDID}(t,S)|),
\end{equation*}
for some constant $C$. Further, as $\Delta_{EDID}\ge 0$,
$\int_0^\infty\int_c^x\te{-qt}|\Delta_{EDID}(t,y)|\td y\td t$ is
for $H<c<x$ equal to $\int_c^x\WH\Delta_{EDID}(t,y)\td y $, which
is finite in view of the form of the delta.  Similarly, it follows
that $\int_0^\infty\int_c^x\te{-qt}|\Theta(t,y)|\td y < \infty$
and the proof is complete.\exit

\noindent Given the sensitivities of $EDID$, those of $ADID$ are
calculated as follows:
\begin{Cor} \label{cor:DDstar}
Let $\s>0$. For $q>0$ and $H<S_0$ it holds that
\begin{eqnarray*}
\Delta_{ADID}(T) &=& \Delta_{EDID}(T) +
r\int_0^T \Delta_{EDID}(s)\td s,\\
\Gamma_{ADID}(T) &=& \Gamma_{EDID}(T) +
r\int_0^T \Gamma_{EDID}(s)\td s.
\end{eqnarray*}
\end{Cor}
\proof{\it of Corollary \ref{cor:DDstar}:} The assertions follow 
the relation between $EDID$ and $ADID$ given in Proposition
\ref{prop:digital}, and using that $\WH f(q)/q$ is the Laplace
transform of $\int_0^T f(s)\td s$. \exit

\noindent Following an analogous approach, similar formulas can be derived
for the sensitivities of down-and-out put options; the results are
reported in Appendix \ref{app:sDIP}.

\section{Numerical results}\label{sec:numerical}
When the stock log-price process is modelled by a generalised
hyper-exponential L\'{e}vy process, we use the following algorithm
to compute single barrier option prices and corresponding
sensitivities:

\bigskip


\medskip

\noindent{\bf Algorithm}

\medskip

{\tt
\begin{enumerate}
\item Approximate the target L\'{e}vy density by a
hyper-exponential density. \item Calculate the Laplace transforms
of prices and sensitivities using \\ the formulas in Sections
\ref{sec:LTbarrier} and \ref{sec:sens}. \item Apply a Laplace
inversion algorithm to the result of step 2.
\end{enumerate}
}

\bigskip \noindent {\bf Comments}

Ad 1. Similarly as in Asmussen et al. \cite{AMP} 
we specified the form of the density
(\ref{eq:ka}) by fixing the number of exponentials
$n^\pm$ and the mean jump sizes $(\a_i^+)^{-1},(\a_j^-)^{-1}$ and
then using a least squares algorithm to determine the values of the 
remaining parameters $p_i^+$, $p_j^-$ and $\lambda_i^+$, $\lambda_j^-$ 
that minimize the squared distance between the 
density (\ref{eq:ka}) and the target density.
To improve the accuracy of the fit one could  
employ methods from the approximation theory of real valued functions,
or alternatively follow the approach developed by 
Feldmann and Whitt \cite{FW}.

Ad 2. The roots of the characteristic exponent that appear in
the formulas of the Laplace transforms were calculated
using Laguerre's method (see e.g. Numerical Recipes \cite{NumRec}).

Ad 3. The quantity of interest $V$ can be expressed in terms of 
the Laplace transform $\WH V(q)$ by the Bromwich integral
\begin{equation}\label{eq:brom}
V(T)=\frac{1}{2\pi \ie }\int_{c-\ie\infty}^{c+\ie\infty} \te{qT}\WH
V(q)dq.
\end{equation}
To evaluate the integral \eqref{eq:brom} we employed the algorithm
investigated by Abate and Whitt \cite{abate95}, which is based on
approximation of the integral \eqref{eq:brom} by
Euler summation with the trapezoidal rule that leads to the series
$$
V(T)\approx\sum_{k=0}^{M} {m\choose k} 2^{-m} s_{N+k}(T),
$$
where $$s_n(T)=\frac{e^{A/2}}{2T}\mbox{Re}\left[\WH V
\left(\frac{A}{2T}\right)\right]+\frac{e^{A/2}}{2T}\sum_{k=1}^n(-1)^k
\mbox{Re}\left[\WH V\left(\frac{A+2k\pi \ie}{2T}\right)\right].$$
Abate and Whitt \cite{abate95} recommend the values $M = 15$,
$N=11$, $A=18.4$, where $N$ should be increased if necessary. See
\cite[p.38-39]{abate95} for a discussion about the role these
parameters play in controlling the error.

We have implemented the above algorithm in the context of a
variance gamma (VG) model and a normal inverse gaussian (NIG)
model to calculate the prices and the sensitivities of digital
barrier options and down-and-out put options. The parameters of the VG
and NIG models were determined by calibrating the models to market
quoted Stoxx50E call prices on 16 June 2006, using the FFT
algorithm proposed by Carr and Madan \cite{CM98}. This procedure
yielded the following parameter values $C=0.925$, $G=4.667$ and
$M=11.876$ for the variance gamma model and  $\alpha=8.858$,
$\beta=-5.808$, $\delta=0.174$ for the normal inverse Gaussian
model. We employed these parameters in the subsequent calculations
of barrier option prices and sensitivities.

As an initial step, the L\'{e}vy densities of the VG and NIG
processes were approximated by hyper-exponential
densities of the form (\ref{eq:ka}) with 14 different exponentials,
7 upward phases ($n^+=7$) and 7 downward phases ($n^-=7$). Note
that the parameters $\a_i^+$ and $\a_j^-$ are equal to the
reciprocals of the mean sizes of upward and downward jumps of the
(hyper-exponential) log-price, respectively. The values of
$\a_i^+$ and $\a_j^-$ were fixed as in Table \ref{table:NIG}. The
remaining parameters $\lambda^+$, $\lambda^-$, and the $p_i^+$,
$p_j^-$ were determined by using a least squares algorithm to
minimize the sums of squares of the difference between the target
density and the hyper-exponential density (\ref{eq:ka}) over a
time-grid inside the interval $[0.1, 5]$. The resulting parameter
values of the hyper-exponential density are specified in Table
\ref{table:NIG}.
\begin{table}[th]
\centering {\small
\begin{tabular}{|c|c|ccccccc|c|c|}
  \hline
\multirow{4}{12mm}{NIG} &  $\alpha^+_i$ & $5$ & $10$ & $15$ & $25$
& $30$ & $60$ & $80$ &
$\lambda^+$ & $5.1$\\
&  $p^+_i$ & $0.005$ & $0.005$ & $0.01$ & $0.06$ & $0.12$ & $0.19$
& $0.61$ & $\sigma^2$ & $0.042$\\ 
\cline{2-11} &  $\alpha^-_j$ & $5$ & $10$ & $15$ & $25$ & $30$ &
$60$ & $80$ &
$\lambda^-$ & $6.4$\\
&  $p^-_j$ & $0.05$ & $0.03$ & $0.11$ & $0.08$ & $0.10$ & $0.40$ &
$0.23$ &$\mu$ & $0.15$\\ 
    \hline
\hline \multirow{4}{12mm}{VG} &  $\alpha^+_i$ & $5$ & $10$ & $15$
& $25$ & $30$ & $60$ & $80$
& $\lambda^+$ & $2.2$ \\
&  $p^+_i$ & $0.003$ & $0.007$ & $0.21$ & $0.08$ & $0.26$ & $0.19$
&
$0.25$ & $\sigma^2$ & 0.011 \\ 
  \cline{2-11}
&  $\alpha^-_j$ & $2$ & $5$ & $10$ & $30$ & $50$ & $80$ & $100$
&$\lambda^-$ & $3$ \\
 & $p^-_j$ & $0.01$ & $0.09$ & $0.31$ & $0.31$ & $0.10$ & $0.08$ & $0.10$
&$\mu$ & 0.13 \\ 
\hline
\end{tabular}
\caption{Parameter values for the approximation of the VG
($C=0.925, G=4.667, M=11.876$) and NIG ($\a=8.858, \beta=-5.808, \d=0.174$)
L\'{e}vy densities} \label{table:NIG} }
\end{table}
\begin{figure}[t]
\centering
    \hspace{-1.3cm}\vspace{-0.2cm}
        \includegraphics[scale =0.44]{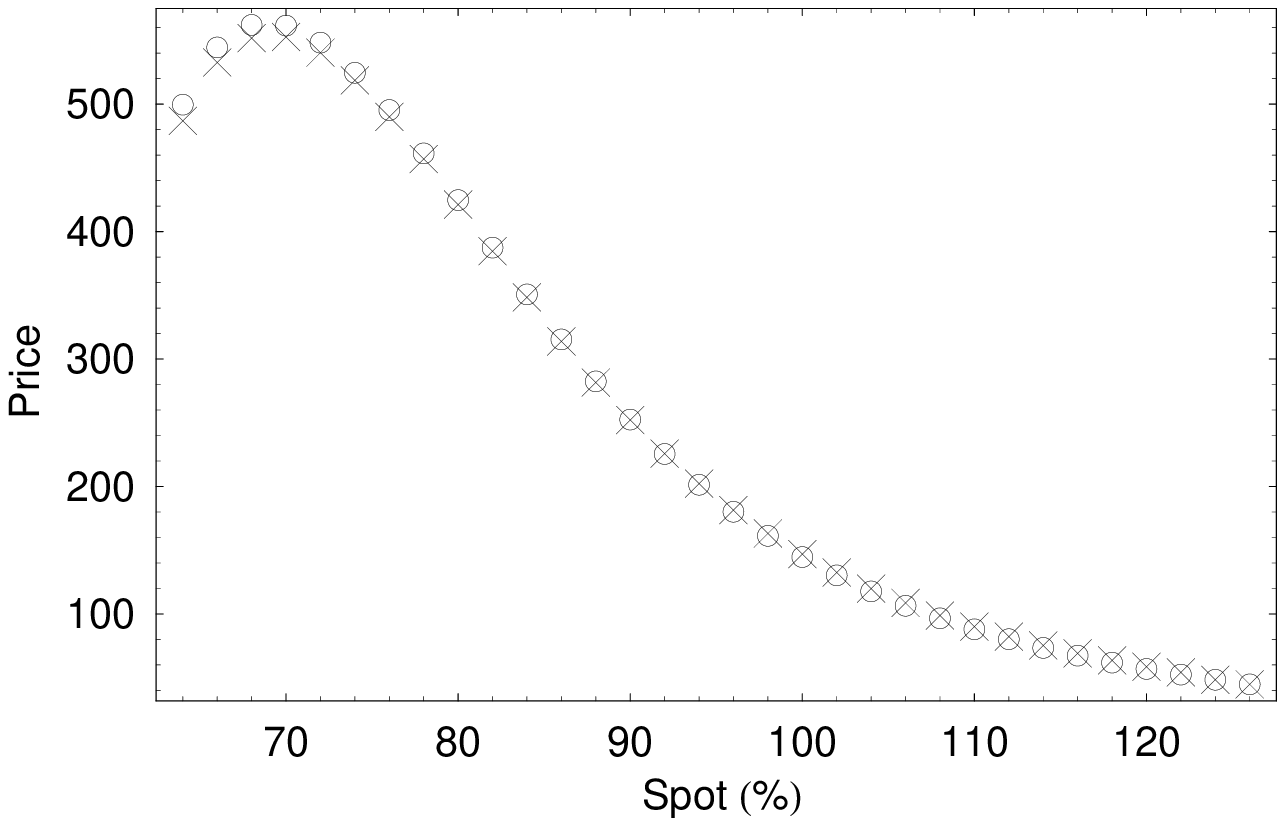}
        \hspace{+1.2cm}
        \includegraphics[scale =0.42]{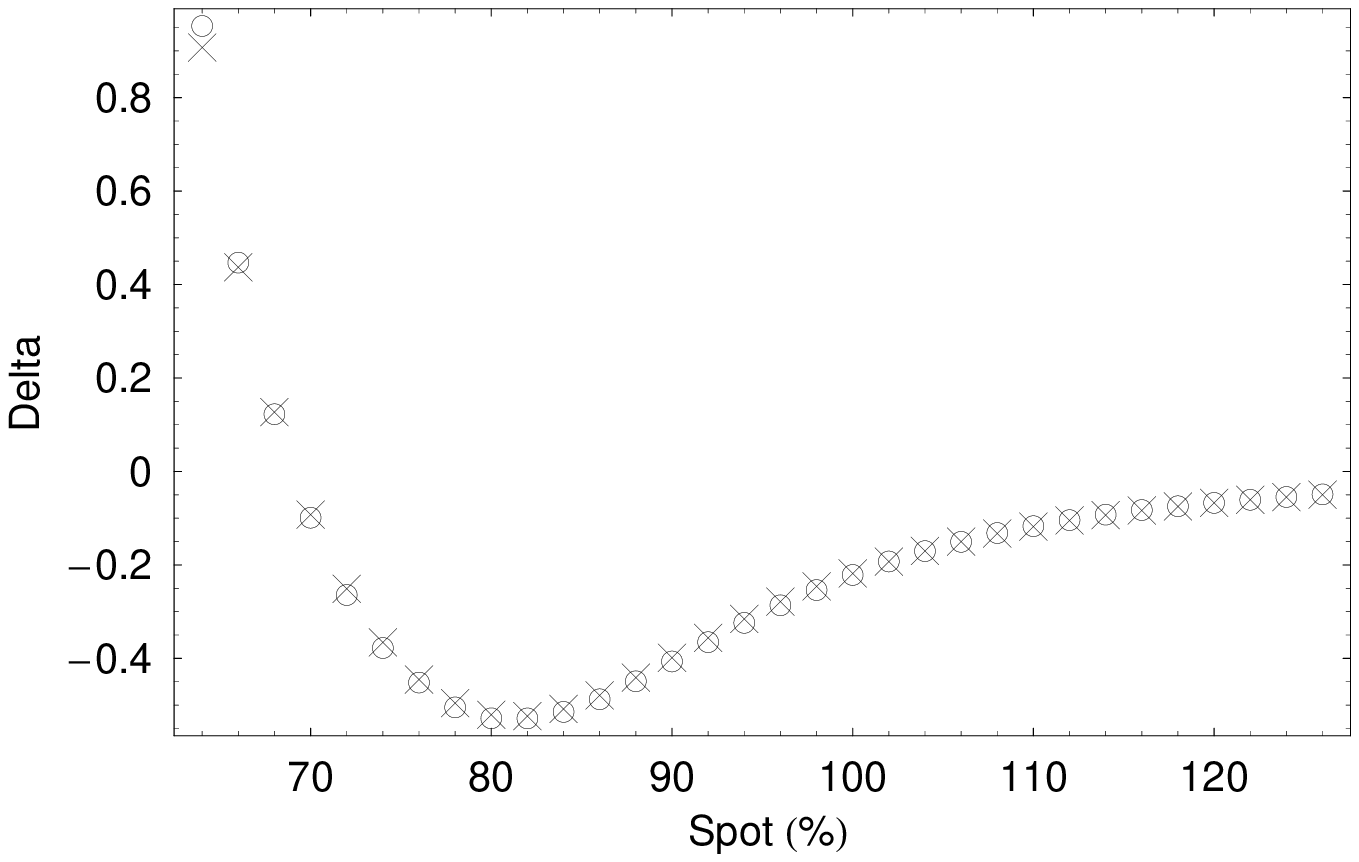}\\

\vspace{+0.5cm} \hspace{-1.8cm}
        \includegraphics[scale =0.42]{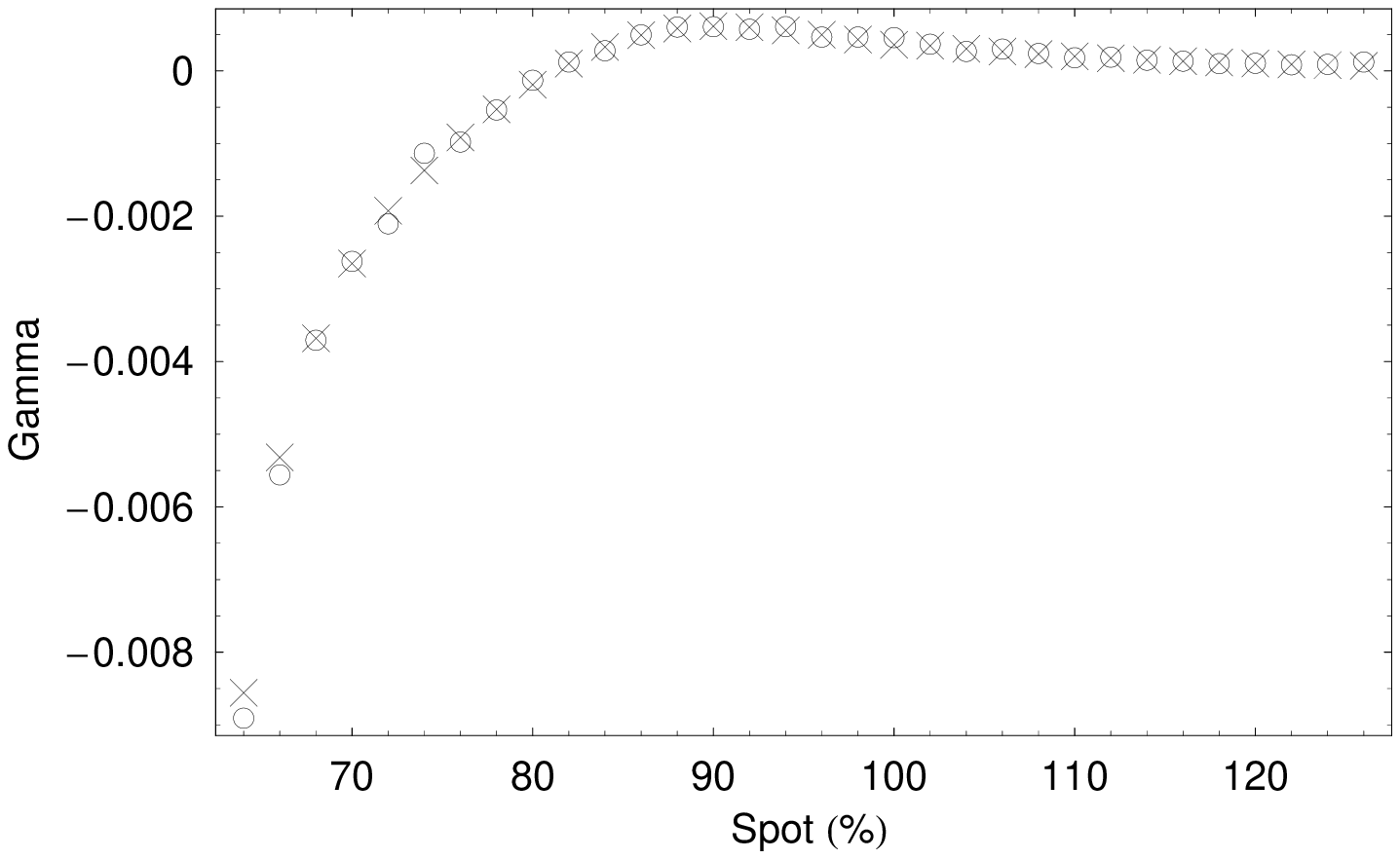}
        \hspace{+1cm}
        \includegraphics[scale =0.42]{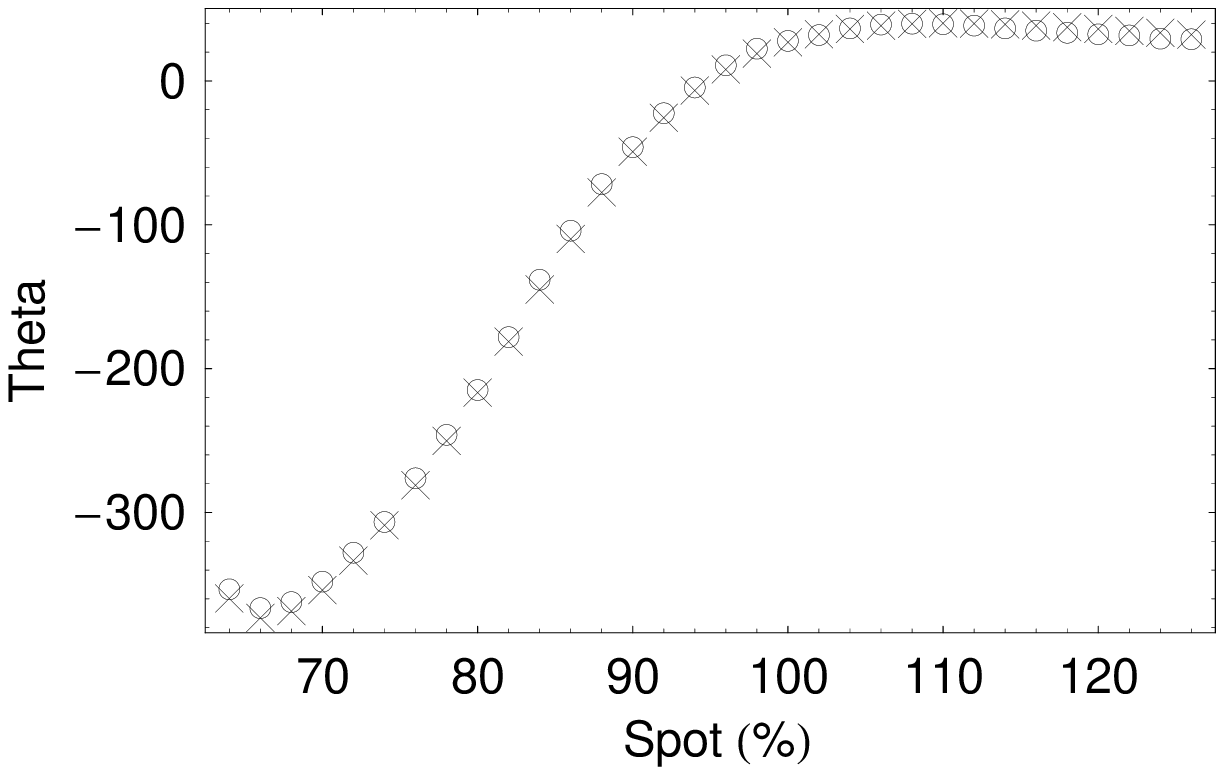}

\caption{\small Prices and sensitivities of a down-and-out put
option with strike $K=3500$ EUR and barrier $H$ set at $60\%$ of
$K$ in an exponential NIG model. Price (left up) - Delta (right
up), Gamma (left-down) and Theta (right down). Results from the
transform algorithm are indicated with a $\times$ and Monte Carlo
results with a $\circ$. } \label{fig:NIG}
\end{figure}

\begin{table}[thp]
\centering {\tiny
\begin{tabular}{|r||rcl|rcl|rc|rc|}
\hline
&\multicolumn{10}{|c|}{Normal inverse Gaussian} \\
&\multicolumn{10}{|c|}{Down-and-out put option} \\
\hline Spot
&\multicolumn{3}{|c|}{Price}&\multicolumn{3}{|c|}{Delta}&
\multicolumn{2}{|c|}{Gamma}&\multicolumn{2}{|c|}{Theta}\\
\hline
    & \multicolumn{1}{|c}{LT}    & \multicolumn{1}{c}{MC} &
    \multicolumn{1}{c|}{RE}&
     \multicolumn{1}{|c}{LT} & \multicolumn{1}{c}{MC}& \multicolumn{1}{c|}{RE}
     & \multicolumn{1}{|c}{LT}  &
 \multicolumn{1}{c}{MC} & \multicolumn{1}{|c}{LT} &
 \multicolumn{1}{c|}{MC} \\
 \hline
    & \multicolumn{1}{|c}{} & \multicolumn{1}{c}{}  &
    \multicolumn{1}{c|}{$\%$}&
\multicolumn{1}{|c}{$\cdot 10$} &
 \multicolumn{1}{c}{$\cdot
10$} & \multicolumn{1}{c|}{$\%$} & \multicolumn{1}{|c}{$\cdot
10^{-3}$} & \multicolumn{1}{c}{$\cdot
10^{-3}$} & \multicolumn{1}{|c}{} & \multicolumn{1}{c|}{}\\
\hline 
64& 486.8&   (493, 507)& 2.7& 9.07&  (9.20, 9.68)&4.1&  -8.56&   (-9.15, -8.98)&   -360& (-365,-342)\\
66& 532.7&   (537, 551)& 2.1& 4.37&  (4.14, 4.61)&0.3&  -5.32&   (-5.47, -5.31)&   -373& (-377,-356)\\
68& 551.8&   (555, 568)& 1.7& 1.27&  (1.07, 1.33)&5.1&  -3.68&   (-3.75, -3.60)&   -369& (-373,-351)\\
70& 552.6&   (555, 567)& 1.5& -0.92& (-1.10, -0.86)&6.9& -2.65&  (-2.65, -2.51)&   -354& (-358,-338)\\
72& 540.4&   (542, 553)& 1.3& -2.51& (-2.70, -2.47)&3.1& -1.93&  (-2.06, -1.92)&   -333& (-338,-318)\\
74& 518.6&   (518, 530)& 1.2& -3.66& (-3.85, -3.62)&2.2& -1.37&  (-1.35, -1.22)&   -309& (-316,-297)\\
76& 490.0&   (490, 501)& 1.1& -4.45& (-4.62, -4.40)&1.4& -0.92&  (-0.95, -0.91)&   -281& (-286,-267)\\
78& 456.9&   (456, 467)& 1.0& -4.96& (-5.13, -4.92)&1.4& -0.53&  (-0.54, -0.50)&   -250& (-255,-237)\\
80& 421.2&   (420, 430)& 0.9& -5.21& (-5.37, -5.16)&1.2& -0.19&  (-0.19, -0.14)&   -217& (-224,-206)\\
82& 384.6&   (382, 393)& 0.8& -5.24& (-5.40, -5.20)&1.3& 0.10&   (0.04, 0.12)&    -181& (-187,-169)\\
84& 348.4&   (346, 356)& 0.6& -5.08& (-5.25, -5.04)&1.3& 0.33&   (0.30, 0.38)&    -145& (-147,-129)\\
86& 313.8&   (310, 321)& 0.5& -4.79& (-4.97, -4.77)&1.7&0.49&   (0.37, 0.49)&    -110& (-112,-95)\\
88& 281.5&   (278, 287)& 0.3& -4.41& (-4.59, -4.39)&2.0& 0.58&   (0.57, 0.69)&    -77.5&  (-80,-63)\\
90& 252.1&   (248, 257)& 0.1& -3.99& (-4.15, -3.96)&1.8& 0.61&   (0.55, 0.67)&    -49.2&  (-54,-38)\\
92& 225.7&   (221, 230)& 0.1& -3.56& (-3.73, -3.54)&2.3& 0.60&   (0.51, 0.63)&    -25.6&  (-30,-14)\\
94& 202.2&   (197, 206)& 0.4& -3.15& (-3.32, -3.15)&2.6& 0.56&   (0.52, 0.64)&    -6.7&   (-12,3)\\
96& 181.5&   (176, 185)& 0.7& -2.79& (-2.94, -2.77)&2.6& 0.50&   (0.44, 0.55)&    8.0&   (3,18)\\
98& 163.1&   (157, 166)& 1.0& -2.46& (-2.59, -2.42)&2.1& 0.43&   (0.43, 0.54)&    18.9&   (15,30)\\
100&    146.9&  (141, 149)&1.3 & -2.18& (-2.27, -2.11)&0.6& 0.36&    (0.36, 0.46)&    26.9& (21,35)\\
102&    132.5&  (127, 135)&1.4 & -1.93& (-2.00, -1.85)&0.1& 0.35&    (0.29, 0.39)&    32.4&  (25,39)\\
104&    119.8&  (114, 122)&1.5 & -1.70&   (-1.78, -1.63)&0.3& 0.30&    (0.25, 0.34)&    36.2&  (30,43)\\
106&    108.7&  (103, 111)&1.7 & -1.50& (-1.57, -1.43)&0.3& 0.27&    (0.23, 0.32)&    38.4&   (33,45)\\
108&    98.76&   (93.5, 101)&1.8 &  -1.33& (-1.39, -1.25)&0.3& 0.23&    (0.19, 0.28)&    39.7&   (34,46)\\
110&    90.00&   (84.9, 91.7)&1.9 &  -1.18& (-1.24,  -1.10)&0.5& 0.20&    (0.14, 0.23)&    40.1&  (34,45)\\
112&    82.22&   (77.3, 83.8)&2.0 &  -1.05& (-1.10, -0.97)&0.8& 0.17&    (0.15, 0.23)&    39.9&   (32,44)\\
114&    75.29&   (70.6, 76.4)&2.1 &  -0.93& (-0.98, -0.86)&1.1& 0.15&    (0.10, 0.17)&    39.3&   (31,42)\\
116&    69.11&   (64.6, 70.6)&2.2 &  -0.84& (-0.89, -0.77)&0.7& 0.13&    (0.10, 0.16)&    38.4&   (30,40)\\
118&    63.57&   (59.2, 65.1)&2.2 &  -0.75& (-0.80, -0.68)&0.7& 0.12&    (0.07, 0.12)&    37.4&   (29,38)\\
120&    58.60&   (54.4,   60.0)&2.4 &  -0.67& (-0.72, -0.61)&0.9& 0.10&    (0.07, 0.11)&    36.1&   (28,37)\\
122&    54.13&   (50.1, 55.5)&2.5 &  -0.61& (-0.65, -0.54)&0.6& 0.09&    (0.04, 0.10)&    34.8&   (27,36)\\
\hline
\end{tabular}
\caption{\small The prices and sensitivities of a down-and-out put
in an exponential NIG L\'{e}vy model with maturity $T= 1$ year,
strike $K=3500$ and barrier $H$ set at $60\%$ of $K$. The interest
and dividend rates are taken to be $r=0.03$ and $d=0$. All columns
except that of the price are expressed in percentage figures of
$K$. The columns with LT contain the results obtained using the
transform algorithm, whereas MC refers to Monte Carlo results,
which are reported in the form of a $95\%$ confidence interval.
The columns with RE contain the relative errors, which is computed
as $|MC^\pm-LT|/LT$ respectively, where $MC^\pm$ is the mid-point
of the confidence interval.} \label{table:NIGDOP} }
\end{table}

\noindent We simulated paths of variance gamma and normal inverse
Gaussian processes by employing the representations of these
processes as random time changes of a Brownian motion by an
independent gamma and an independent 
inverse Gaussian subordinator, respectively. The
sensitivities were calculated by approximating the derivative
by a finite difference and subsequently evaluating this using
Monte Carlo simulation.

\begin{figure}[t]
\centering
    \hspace{-1.3cm}
    \vspace{-0.2cm}
        \includegraphics[scale =0.42]{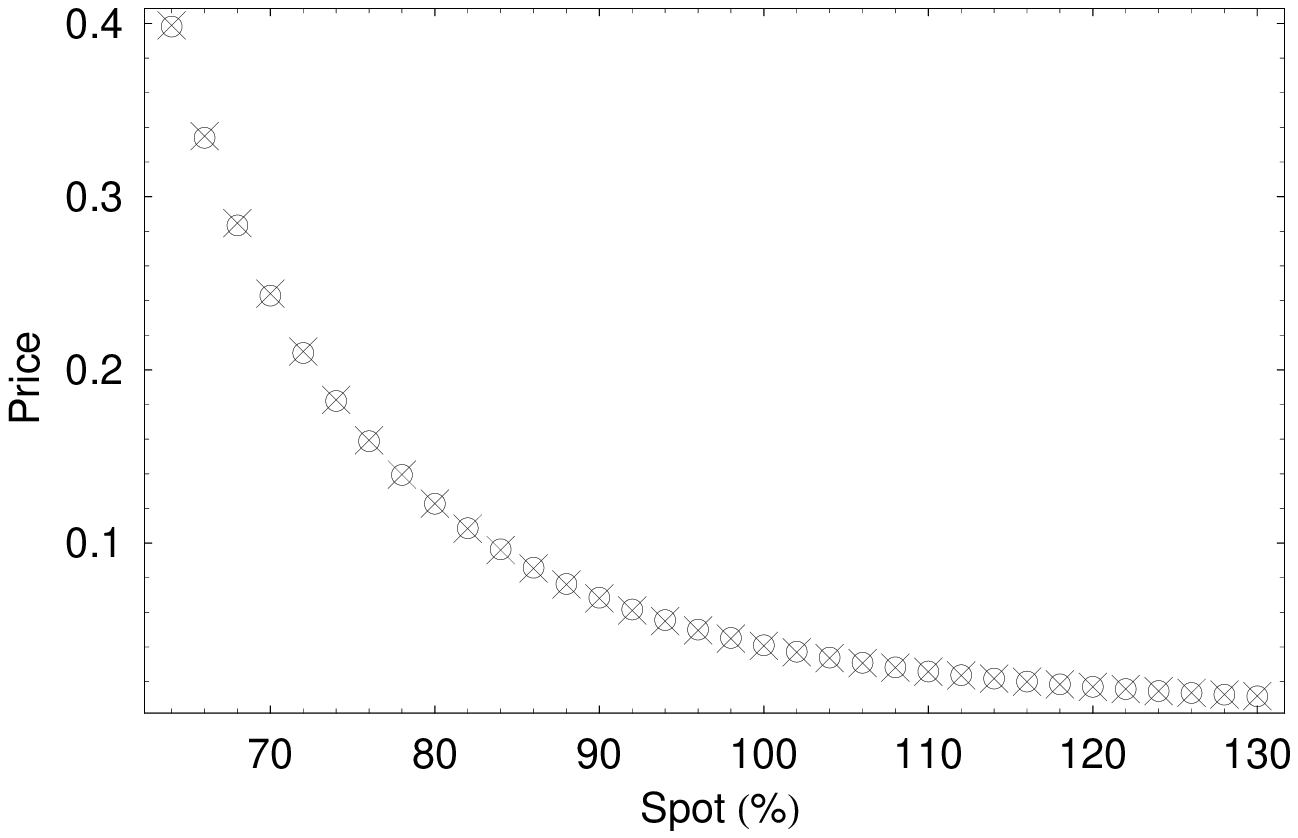}
        \hspace{+0.7cm}
        \includegraphics[scale =0.42]{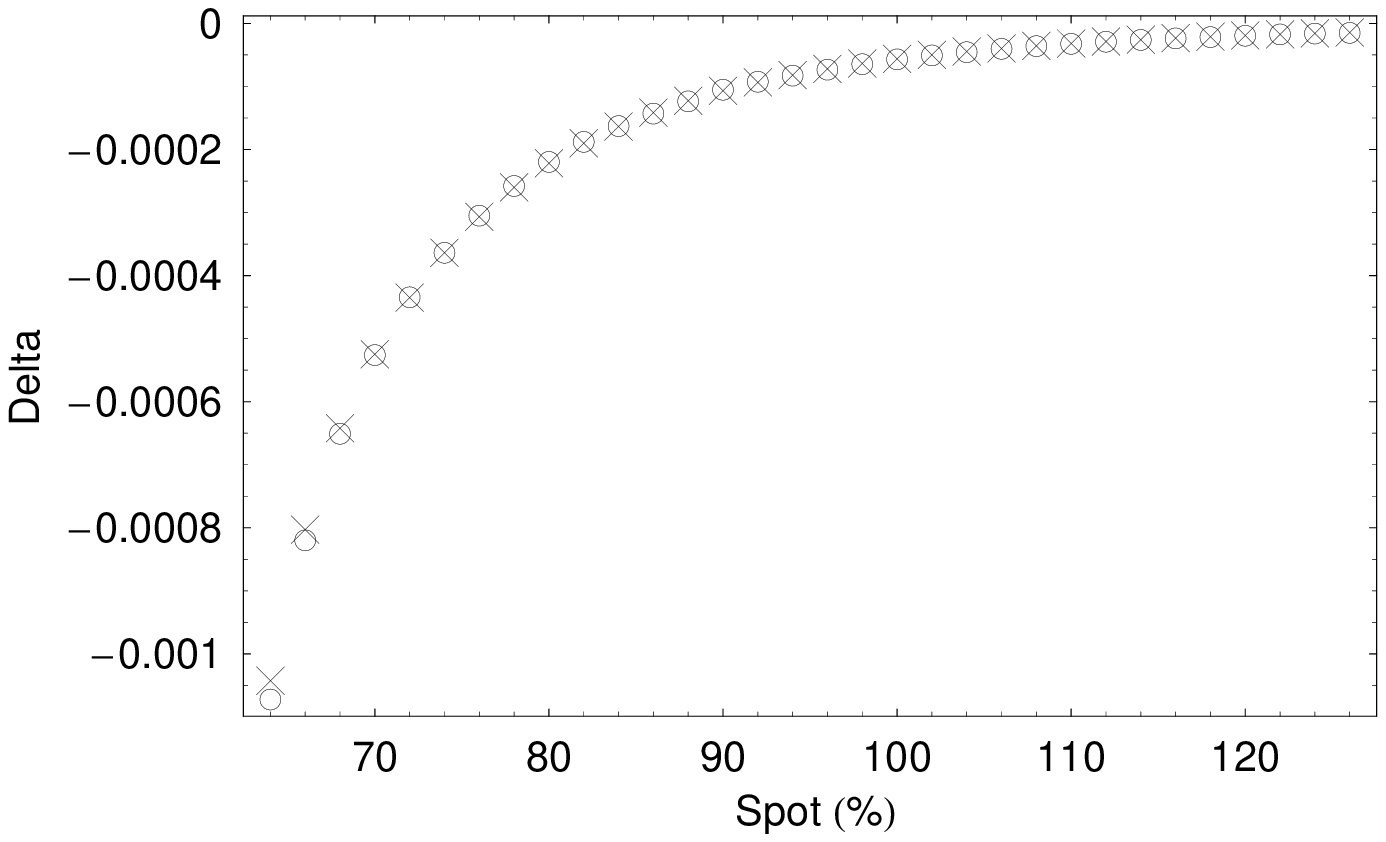}\\
                \vspace{+0.5cm}\hspace{-1.3cm}
               \includegraphics[scale =0.42]{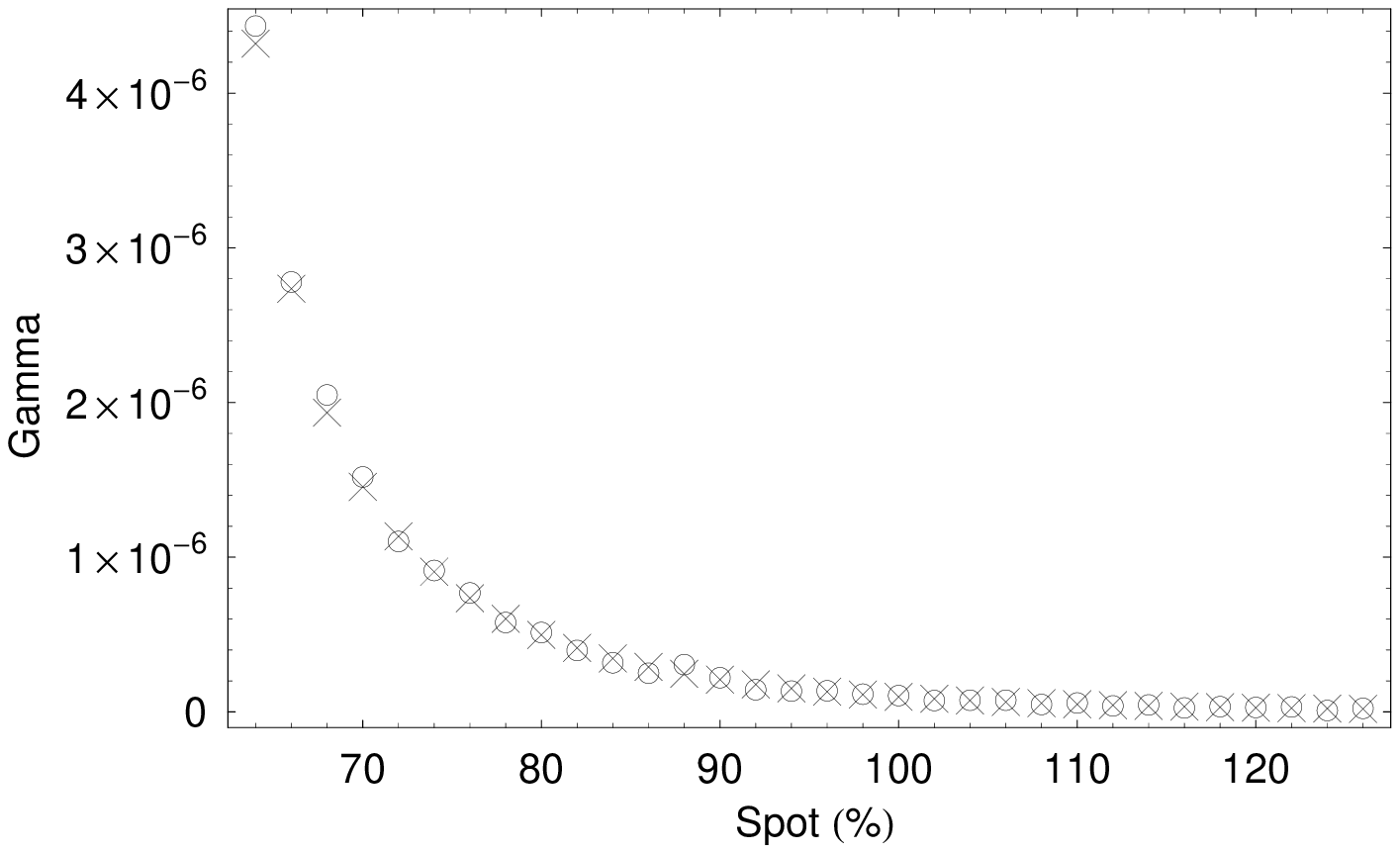}
        \hspace{+0.7cm}
        \includegraphics[scale =0.42]{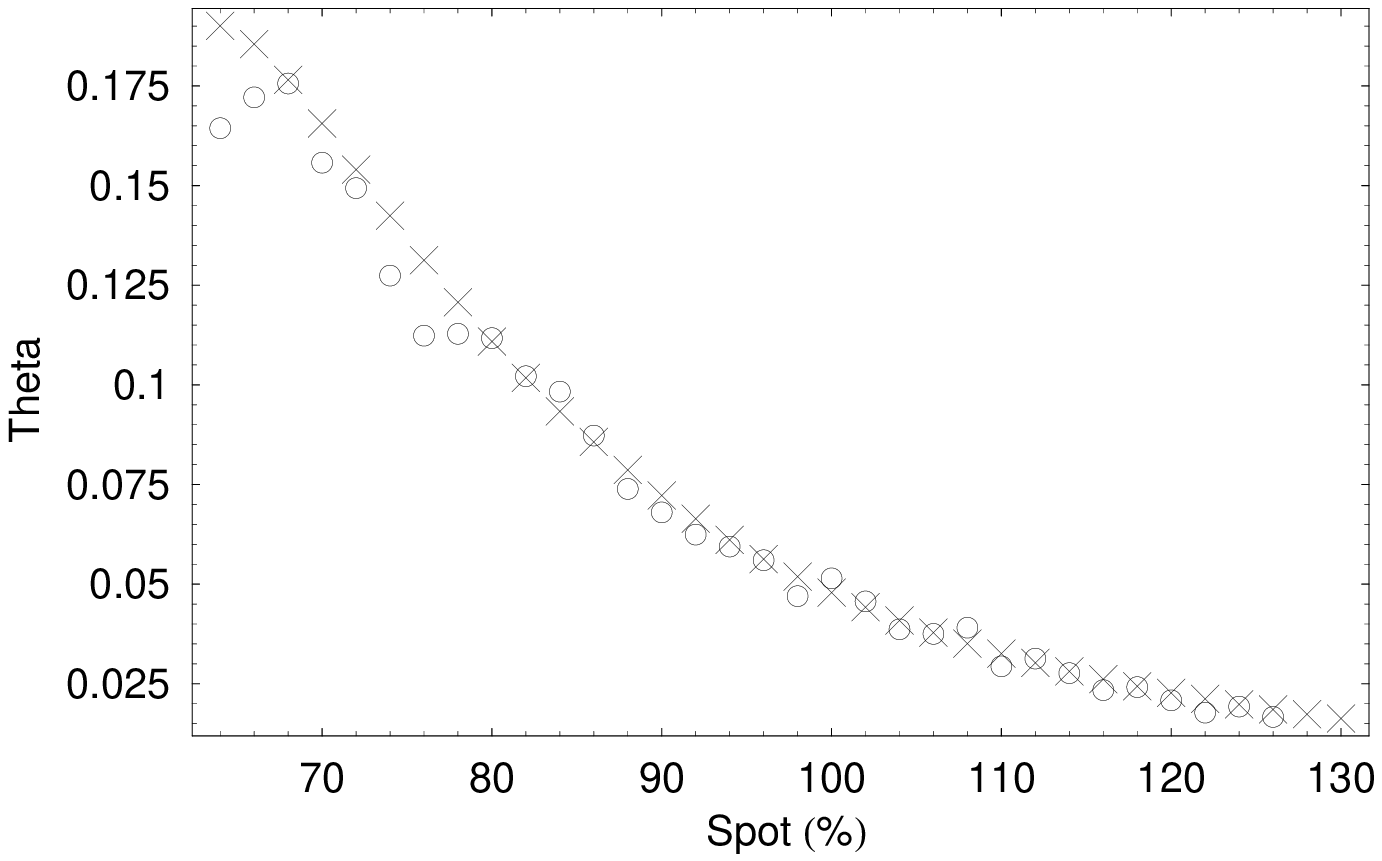}
\caption{ {\small Prices and sensitivities of an American
down-and-in digital option with a barrier $H$ set at $60\%$ of EUR
$3500$ in an exponential VG model. Price (left up) - Delta (right
up), Gamma (left-down) and Theta (right down). Results from the
transform algorithm are indicated with a $\times$ and Monte Carlo
results with a $\circ$.} } \label{fig:VG}
\end{figure}

Tables \ref{table:NIGDOP} and \ref{table:VGDD} and Figures
\ref{fig:NIG} and \ref{fig:VG} report prices and greeks for
down-and-in digital and down-and-out put options calculated by
using Algorithm 1 and Monte-Carlo simulation. The options are
priced at different spot levels that are expressed as a percentage
of the spot price on June 16th 2006 ($S_0=3500$). The value of the
sensitivities is expressed as a fraction of $S_0$. Down-and-in
digital options are priced under the VG model with barrier level
$H$ set at $60\%$, and down-and-out puts under the NIG model with
a barrier $H$ and a strike $K$ set respectively at $60\%$ and
$100\%$.

The results show a general agreement between the Monte Carlo
simulation results and those from the semi-analytical
approximation. The relative errors for the prices are less than
0.6\% for VG and about 2\% for NIG, and for the delta about 2\% -
2.5\% for both, if, for NIG, we stay some distance away from the
barrier. We also observe that, close to the barrier, the result
for the gamma falls outside the confidence interval. In the case
of the NIG L\'{e}vy model, the larger errors close to the barrier
may be explained by the fact that the option price is not smooth
under this model whereas it is for the approximating
hyper-exponential model (see \cite{BoyarchenkoBook} for an
analysis of the smoothness of option prices under an NIG or, more
generally, a regular L\'{e}vy process).

The convergence of Monte Carlo estimators of quantities involving
first passage is known to be very slow, requiring a small mesh of
the time grid. Since this effect is magnified when calculating the
numerical derivatives, a large number of paths and time-steps was
needed to obtain convergence. For the Monte Carlo calculations we
used $1,000,000$ paths and $20,000$ time steps per year, which
made the calculation of the greeks using Monte Carlo
computationally demanding and time-consuming. In a C++ programme
on a 3189 Mhz computer the valuation of all option prices and
sensitivities took $40$ seconds for a digital option and $55$
seconds for a down-and-out put option, compared to a couple of
hours when using Monte Carlo. Here we should remark that we did
not employ any special methods that could considerably have
improved the speed of convergence of the Monte Carlo simulation,
such as Malliavin weights, likelihood ratio methods or sampling
techniques such as the bridge (see e.g. Glasserman \cite{glas}).
We noted that if the spot price was close to the barrier the Monte
Carlo estimators of the greeks tended to be unstable.

\begin{table}[thp]
\centering {\tiny
\begin{tabular}{|r||rcl|rcl|rc|rc|}
\hline
&\multicolumn{10}{|c|}{Variance Gamma} \\
&\multicolumn{10}{|c|}{Down-and-in digital option - payment at first passage} \\
\hline Spot
&\multicolumn{3}{|c|}{Price}&\multicolumn{3}{|c|}{Delta}&
\multicolumn{2}{|c|}{Gamma}&\multicolumn{2}{|c|}{Theta}\\
\hline
    & \multicolumn{1}{|c}{LT}    & \multicolumn{1}{c}{MC} &
    \multicolumn{1}{c|}{RE}&
     \multicolumn{1}{|c}{LT} & \multicolumn{1}{c}{MC}& \multicolumn{1}{c|}{RE}
     & \multicolumn{1}{|c}{LT}  &
 \multicolumn{1}{c}{MC} & \multicolumn{1}{|c}{LT} &
 \multicolumn{1}{c|}{MC} \\
\hline
    & \multicolumn{3}{|r|}{$\cdot 10^{-2}$ \qquad  $\cdot 10^{-2}$\quad $\cdot 10^{-3}$} &
\multicolumn{3}{|r|}{$\cdot 10^{-5}$ \qquad\quad $\cdot 10^{-5}$
\qquad $\%$}& \multicolumn{1}{|c}{$\cdot 10^{-7}$}&
\multicolumn{1}{c|}{$\cdot 10^{-7}$}&
\multicolumn{1}{|c}{$\cdot10^{-2}$} &
\multicolumn{1}{c|}{$\cdot 10^{-2}$}\\
\hline 64& 39.89&(39.6, 40.1)&0.5&-104&  (-108, -106)&2.9&
43.2&(44.4, 45.4)&19.0&(16.3,19.4) \\
66& 33.50&(33.2, 33.7)&1.0&-80.3&  (-82.7, -81.0)&1.9&
27.4&(27.4,
28.3)&18.5& (16.3,19.3)\\
68& 28.47&(28.2, 28.6)&2.1&-64.2&  (-65.7, -64.3)&1.3&
19.4&(19.5,
20.9)&17.6& (16.2,19.1)\\
70& 24.41&(24.1, 24.6)&2.2&-52.5&  (-53.4, -52.1)&0.7&
14.6&(14.1,
15.2)&16.6&(14.8,17.6)\\
72& 21.06&(20.8, 21.2)&2.4&-43.5&  (-44.2, -43.0)&0.3&
11.4&(10.9,
12.2)&15.4& (13.2,15.8)\\
74& 18.27&(18.1, 18.4)&2.1&-36.4&  (-36.9, -35.8)&0.0&
9.06&(8.55,
9.57)&14.2& (12.5,15.0)\\
76& 15.93&(15.8, 16.1)&1.5&-30.7&  (-31.0, -30.1)&0.4&
7.34&(7.07,
7.99)&13.1& (11.6,14.0)\\
78& 14.96&(13.8, 14.1)&0.3&-26.0&  (-26.3, -25.4)&0.4&
6.02&(5.35,
6.17)&12.1& (11.2,13.5)\\
80& 12.27&(12.2, 12.4)&0.2&-22.2&  (-22.3, -21.6)&0.8&
4.97&(4.85,
5.89)&11.1& (10.1,12.2)\\
82& 10.84&(10.8, 11.0)&2.6&-19.0&  (-19.1, -18.4)&0.9&
4.13&(3.22,
4.15)&10.2& (9.4,11.4)\\
84& 9.60&(9.53, 9.72)&3.3&-16.4&  (-16.6, -16.0)&0.2&  3.46&(3.05, 3.89)
&9.34& (8.2,10.2)\\
86& 8.54&(8.48, 8.65)&4.2&-14.1&  (-14.4, -13.8)&0.1&  2.90&(2.36,
3.11)&8.57& (7.6,9.4)\\
88& 7.61&(7.56, 7.72)&4.0&-12.3&  (-12.5, -12.0)&0.5&  2.45&(2.14,
2.82)&7.87&(6.7,8.4)\\
90& 6.81&(6.76, 6.90)&4.0&-10.7&  (-10.8, -10.4)&0.4&  2.08&(2.06,
2.64)&7.22& (6.6,8.2)\\
92& 6.11&(6.08, 6.21)&5.9&-9.3&  (-9.4, -9.1)&0.6&  1.77&(1.26,
1.81)&6.64& (6.2,7.8)\\
94& 5.50&(5.47, 5.58)&6.0&-8.2&  (-8.4, -8.0)&0.1&  1.52&(1.28,
1.79)&6.11&(5.9,7.4)\\
96& 4.96&(4.94, 5.04)&6.3&-7.2&  (-7.4, -7.1)&0.7&  1.30&(1.08,
1.38)&5.63&(5.4,6.8)\\
98& 4.49&(4.46, 4.55)&4.9&-6.4&  (-6.5, -6.2)&0.6&  1.12&(0.92,
1.29)&5.19&(4.9,6.3)\\
100&4.07&(4.04, 4.13)&5.2&-5.6&  (-5.8, -5.5)&0.6&  0.97&(0.64, 0.99)&4.79& (4.3,5.2)\\
102&3.70&(3.66, 3.74)&3.4&-5.0&  (-5.1, -4.9)&1.0&  0.84&(0.54, 0.89)&4.42& (4.1,5.3)\\
104&3.37&(3.25, 3.53)&2.8&-4.5&  (-4.6, -4.4)&0.3&  0.73&(0.47, 0.84)&4.09& (3.6,4.8)\\
106&3.07&(3.04, 3.11)&1.8&-4.0&  (-4.1, -3.9)&0.1&  0.64&(0.40, 0.78)&3.78& (3.5,4.6)\\
108&2.81&(2.78, 2.84)&1.2&-3.6&  (-3.7, -3.5)&1.4&  0.56&(0.31, 0.72)&3.51& (3.2,4.3)\\
110&2.57&(2.55, 2.60)&2.6&-3.2&  (-3.3, -3.1)&1.4&  0.49&(0.28, 0.63)&3.25& (3.0,4.0)\\
112&2.36&(2.33, 2.38)&1.5&-2.9&  (-3.0, -2.8)&0.3&  0.43&(0.25, 0.51)&3.02& (2.7,3.7)\\
114&2.17&(2.14, 2.19)&0.9&-2.6&  (-2.7, -2.5)&0.2&  0.38&(0.23, 0.48)&2.81& (2.5,3.4)\\
116&2.00&(1.98, 2.02)&1.8&-2.3&  (-2.4, -2.2)&1.8&  0.34&(0.22, 0.41)&2.62& (2.3,3.2)\\
118&1.84&(1.83, 1.86)&0.8&-2.1&  (-2.2, -2.1)&0.2&  0.30&(0.21, 0.37)&2.44& (2.2,3.1)\\
120&1.70&(1.68, 1.71)&1.3&-1.9&  (-2.0, -1.9)&0.5&  0.27&(0.17, 0.33)&2.27& (2.0,2.9)\\
122&1.58&(1.57, 1.59)&4.0&-1.7&  (-1.7, -1.7)&0.6&  0.24&(0.16, 0.30)&2.12& (1.8,2.6)\\
124&1.46&(1.45, 1.47)&4.9&-1.6&  (-1.6, -1.6)&2.3&  0.21&(0.15, 0.26)&1.98& (1.7,2.5)\\
126&1.35&(1.34, 1.36)&5.8&-1.4&  (-1.5, -1.4)&2.2&  0.19&(0.14, 0.23)&1.86& (1.6,2.4)\\
\hline
\end{tabular}
\caption{\small The prices and sensitivities of an American
down-and-in digital option with maturity $T= 1$ year barrier and
$H$ set at $60\%$ of $3500$ in an exponential VG L\'{e}vy model.
The interest and dividend rates are taken to be $r=0.03$ and
$d=0$. All columns except that of the prices are expressed in
percentage figures of $3500$. The columns with LT contain the
results obtained using the transform algorithm, whereas MC refers
to Monte Carlo results, which are reported in the form of a $95\%$
confidence interval. The columns RE contain the relative error,
which is computed as $|MC^\pm-LT|/LT$ respectively, where $MC^\pm$
is the mid-point of the confidence interval.\label{table:VGDD}} }
\end{table}

\section{Convergence of prices and sensitivities}\label{sec:convsens}
A generalised hyper-exponential L\'{e}vy process can, as we have shown in Section \ref{sec:genhyp}, be approximated arbitrarily closely by a hyper-exponential jump-diffusion, by specifying appropriate values for the parameters 
of the latter process.
Under such a model analytical expressions were derived in 
Sections \ref{sec:LTbarrier} and \ref{sec:sens} for the values  
$EDID$, $EDOD$, $ADID$, $DOP$ of digital and the down-and-out put options, and the corresponding sensitivities. 
The results in Section \ref{sec:numerical} 
provided numerical evidence to show that a good approximation of prices and sensitivities in a $GHE$ L\'{e}vy model can be obtained by carrying out the computations in a suitably chosen approximating hyper-exponential jump-diffusion model. In this section we provide further justification for the algorithm by proving a convergence result, that is, for a given 
$GHE$ L\'{e}vy process $X$ and a 
sequence of $HEJD$ processes $X^{(n)}$ weakly converging to $X$ 
(as constructed in Section \ref{sec:genhyp}), we will show that the 
corresponding prices $EDID^{(n)}$, $EDOD^{(n)}$,
$ADID^{(n)}$, $DOP^{(n)}$ of digital and down-and-out put options converge pointwise to those in the limiting model $X$, as $n\to\infty$:

\begin{Prop}\label{prop:convbar}
Suppose that $X$ is $GHE$ and is not of type A and let $V$ be any
of $EDID, ADID, EDOD, DOP$ with $V^{(n)}$ the corresponding
approximation. Then for $S_0>H, T>0$, it holds that
\begin{equation}\label{conv:price}
V^{(n)}(T,S_0) \to V(T,S_0) \quad \text{as $n\to\infty$.}
\end{equation}
\end{Prop}

\proof{\it of Proposition \ref{prop:convbar}} \
The convergence of $EDID^{(n)}(T,S_0)$ to $EDID(T,S_0)$ is
a direct consequence of the convergence
of $P(T^{(n)}(h) < t)$ (Proposition \ref{prop:convWH}).
Further, by interchanging the order of integration
it follows that the value function $ADID$ can be written as
$$ADID(T) = \int_0^\infty r\te{-rt}P(T(h) < T\wedge t)\td t.$$
Thus, in view of the dominated convergence theorem and again 
Proposition \ref{prop:convWH} it follows that
$ADID^{(n)}(T,S_0)\to ADID(T,S_0)$. Also, since $(K-\te{x})^+$
is a continuous bounded function, Lemma \ref{lem:weakXX} implies
that $E[(K-\te{X^{(n)}(T)})^+\mbf 1_{\{T^{(n)}(h)<T\}}] = DOP^{(n)}(T;S_0)$
converges to $DOP(T;S_0)$ as $n\to\infty$. \exit
\medskip

Note that Proposition \ref{prop:convbar} implies that the finite
difference approximations of the sensitivities of the digital and
down-and-out put options also converge pointwise as $n\to\infty$ 
for any spot price $S_0$ away from the barrier $H$ and any maturity $T>0$. To
rigorously prove pointwise convergence of the sensitivities,
uniform estimates would be needed of the errors in
\eqref{conv:price}. It is worth noting that, while for L\'{e}vy
processes with positive Gaussian component first-passage
probabilities and value-functions of barrier options are smooth up
to the barrier, such is not necessarily the case for a L\'{e}vy
process without Gaussian component, especially at the barrier
where the spatial derivatives may be infinite. However, 
when suitably {\it smoothed}, the sensitivities do converge pointwise:

\begin{Cor}\label{cor:sens} For some $\e>0$ let
$V^\e(s)=\int V(T,y)\phi_\e(s-y)\td y$, where $\phi_\e$ is a $C^2$
probability density with support in $(-\e,\e)$. Then it holds that
$$
\Delta_{V_\e^{(n)}}(S_0,T) \to \Delta_{V_\e}(S_0,T), \quad
\Gamma_{V_\e^{(n)}}(S_0,T) \to \Gamma_{V_\e}(S_0,T),
$$
for $S_0>H+\e, T>0$, as $n\to\infty$.
\end{Cor}
A similar statement holds true for the theta, replacing smoothing
in space by smoothing in time. Note that, if $s\mapsto V(T,s)$ is
$C^1$ on $(H,\infty)$, then
$\Delta_{V_\e}(S_0,T)-\Delta_V(S_0,T)$, $S_0>H$, can be made
arbitrarily small by choosing $\e$ small enough (similar remarks
apply to the other greeks)

\proof
In view of the definition
of $V^{(n)}_\e$ and integration by parts it follows that
\begin{equation}\label{eq:deltavn}
\Delta_{V^{(n)}_\e}(S_0,T) = \frac{\partial}{\partial S_0}
{V^{(n)}_\e}(S_0,T) = \int V^{(n)}(y) \phi_\e'(S_0-y)\td y.
\end{equation}
By the dominated convergence theorem and Proposition
\ref{prop:convbar} it thus follows that the rhs of \eqref{eq:deltavn} 
converges to  $\int
V(y) \phi_\e'(S_0-y)\td y = \Delta_{V_\e}(S_0,T)$ as $n\to\infty$.
The convergence of the gamma follows by a similar reasoning.\exit

\section{Conclusion}
In this paper we developed an efficient algorithm to compute
prices and sensitivities of barrier options driven by an
exponential L\'{e}vy model of generalised hyper-exponential type.
We showed that the latter class contains many of the L\'{e}vy
models that are employed in mathematical finance. We first
approximated the target L\'{e}vy measure by a hyper-exponential
one. Subsequently, for log-price processes in this class,
jump-diffusions with hyper-exponential jumps, we derived
analytical expressions for the prices and sensitivities (greeks) of 
digital, knock-in and knock-out option prices, 
up to a single Laplace transform. Inversion of this Laplace transform 
yielded fast and accurate
results for the option prices and sensitivities. We proved
convergence of this algorithm. To provide a numerical illustration 
we implemented the algorithm for
the VG and NIG L\'{e}vy models,approximating the NIG and VG L\'{e}vy densities by hyper-exponential ones with 7 upward and 7 downward
phases. Compared with Monte Carlo simulation results we found
relative errors of about 0.5-2.5\% for prices and deltas some
distance away from the barrier. 
What the rate of convergence of this algorithm is when increasing the number of terms in the approximation, and how this rate depends on the quality of the approximation of the target density, the parameters and 
the distance to the barrier, are open questions that are left for
future research.

\newpage
\appendix
\centerline{\bf APPENDIX}

\section{Additional pricing formulas}\label{app:DIP}
\subsection{Upward digital options}
The arbitrage free prices of the corresponding
up-and-in digitals (contracts that pays out $\pounds 1$ if an upper
barrier is crossed, either at $T$ or at the moment $T^+(h)$
of up-crossing) are given by
$$EUID(T,S_0,H) = \te{-rT}P[\ovl S(T) > H],
\quad AUID(T,S_0,H) = E[\te{-rT^+(h)}\mbf 1_{\{\ovl S(T) >
H\}}],
$$
where $\ovl S(T)=\sup_{0\leq s\leq T}S(s)$ denotes the running
supremum of $S$ up to $T$. Their respective Laplace transforms in
$T$ follow by applying the formulas of the down-and-in digital
options to the process $-X$:
\begin{equation*}
\WH{EUID}(q) = \frac{1}{q+r}\sum_{i=1}^{k^+}A_i^+\te{-\rho_i^+h},\quad
\WH{AUID}(q) = \frac{q+r}{q} \WH{EUID}(q),\quad h\ge0.
\end{equation*}
\subsection{Down-and-in put}
The Laplace transform $\WH{DIP}(q)$ of the arbitrage free price
of the down-and-in put option $DIP(T;S_0,K,H)$ with strike $K$
and barrier $H$ can be decomposed as
\begin{eqnarray*}
\WH{DIP}(q) &=& \frac{K}{q+r} P[\unl X(\t) < h, X(\t)<k] -
\frac{S_0}{q+r}E[\te{X(\t)}\mbf 1_{\{\unl X(\t) < h, X(\t)<k\}}]\\
&=& \frac{K}{q+r}D^{(0)}(h,k) - \frac{S_0}{q+r}D^{(1)}(h,k),
\end{eqnarray*}
where $h=\log (H/S_0)$, $k=\log(K/S_0)$ and
$D^{(b)}(h,k)$ is defined as
$$
D^{(b)}(h,k) := E[\te{bX(\t)}\mbf 1_{\{\unl X(\t) < h,
X(\t)<k\}}].
$$
Reasoning as in Proposition \ref{prop:barrier} we find that
\begin{eqnarray*}
D^{(b)}(h,k) &=& \int_{-h}^\infty\te{-by} E[\te{b(X(\t)-\unl
X(\t))} \mbf 1_{\{X(\t)-\unl X(\t)<k+y\}}]P(-\unl X(\t)\in\td y).
\end{eqnarray*}
After some calculations we arrive at
\begin{eqnarray*}
D^{(b)}(h,k) &=& \sum_{j=1}^{k^-}\sum_{i=1}^{k^+}
\frac{\rho_i^+\rho_j^-}{(\rho_j^--\rho_i^+)(b-\rho_i^+)}A_i^+A_j^-\te{(b-\rho_i^+)k+(\rho_i^+-\rho_j^-)h}\\
&+& \le(1 - \sum_{i=1}^{k^+}\frac{b}{b-\rho_i^+}A_i^+\ri)
\sum_{j=1}^{k^-}A_j^-\frac{\rho_j^-}{\rho_j^--b}\te{(b-\rho_j^-)h}.
\end{eqnarray*}

\section{Sensitivities for down-and-out-puts}\label{app:sDIP}

\begin{Prop}\label{prop:sdop1}
Suppose $X$ is a hyper-exponential L\'{e}vy process
with $\s>0$ and let $q>0$ and $H<\min\{K,S_0\}$.

(i) It holds that $\WH\Th_{DOP}(q) = q \WH{DOP}(q) - (K-S_0)^+$.

(ii) If $S_0<K$, it holds that
\begin{eqnarray*}
\WH\Delta_{DOP}(q) &=& \sum_{j=1}^{k^-}A_j^-(\rho_j^-)^2 S_0^{\rho_j^- - 1}
[KB^{(0)}_{j} - B^{(1)}_j],
\\
\WH\Gamma_{DOP}(q) &=& \sum_{j=1}^{k^-}A_j^-(\rho_j^-)^2(\rho_j^--1) S_0^{\rho_j^- - 2}
[KB^{(0)}_{j} - B^{(1)}_j],
\end{eqnarray*}
where
\begin{eqnarray*}
B^{(b)}_j &=& \sum_{i=1}^{k^+}A_i^+\le[
\frac{(-\rho_i^+)H^{\rho_i^+-\rho_j^-}K^{b-\rho_i^+}}{(\rho_j^--\rho_i^+)(b-\rho_i^+)}
+ \frac{bH^{b-\rho_i^+}}{(b-\rho_j^-)(b-\rho_i^+)} -
\frac{(-\rho_i^+)K^{b-\rho_j^-}}{(\rho_j^--\rho_i^+)(b-\rho_j^-)}\ri]\\
&+& \frac{1}{b-\rho_j^-}[H^{b-\rho_j^-} - K^{b-\rho_j^-}].
\end{eqnarray*}
(iii) If $K<S_0$, it holds that
\begin{eqnarray*}
\WH\Delta_{DOP}(q) &=& \sum_{j=1}^{k^-}A_j^-(\rho_j^-)^2 
S_0^{\rho_j^- - 1}
[KF^{(0)}_{j} - F^{(1)}_j] \\ 
&\phantom{=}& + \sum_{i=1}^{k^+}A_i^+(\rho_i^+)^2 S_0^{\rho_i^+ - 1}
[KG^{(0)}_{i} - G^{(1)}_i] + \d_+\d_-,
\\
\WH\Gamma_{DOP}(q) &=& \sum_{j=1}^{k^-}A_j^-(\rho_j^-)^2(\rho_j^--1) S_0^{\rho_j^- - 2}
[KF^{(0)}_{j} - F^{(1)}_j] \\ &\phantom{=}& + 
\sum_{i=1}^{k^+}A_i^+(\rho_i^+)^2(\rho_i^+-1) S_0^{\rho_i^+ - 2}
[KG^{(0)}_{j} - G^{(1)}_j],
\end{eqnarray*}
where $\d_+ = 1 - \sum_{i=1}^{k^+}A_i^+\frac{1}{1-\rho_i^+}$ and
$\d_- = 1 - \sum_{j=1}^{k^-}A_j^-\frac{1}{1-\rho_j^-}$ and
\begin{eqnarray*}
F^{(b)}_j &=& \frac{H^{b-\rho_j^-}\rho_j^-}{b-\rho_j^-} + \sum_{i=1}^{k^+}A_i^+
\le[\frac{(-\rho_i^+)K^{b-\rho_i^+}H^{\rho_i^+-\rho_j^-}}{(\rho_j^--\rho_i^+)(b-\rho_i^+)}
- \frac{H^{b-\rho_j^-}b}{(b-\rho_j^-)(b-\rho_i^+)}\ri],\\
G^{(b)}_i &=& \frac{K^{b-\rho_i^+}}{b-\rho_i^+}\le[
\sum_{j=1}^{k^-}A_j^-\frac{\rho_i^+}{\rho_j^--\rho_i^+} + 1 \ri].
\end{eqnarray*}
\end{Prop}

\proof{\it of Proposition \ref{prop:sdop1}.}
The formula of the Laplace transform of the theta is a
consequence of integration by parts and the fact that
$G(\cdot,h,k)$ is continuously differentiable
on $(0,\infty)$. Those of the Laplace transforms of
the delta and the gamma follow by differentiating
the expressions in Proposition \ref{prop:barrier},
following a reasoning analogous to the one 
in Proposition \ref{prop:DDstar}.\exit

\newpage
\section{Proofs of the weak convergence}\label{sec:weak}

Let $X$ be a generalised hyper-exponential L\'{e}vy process
and define $X^{(n)}$ as in Section
\ref{sec:WH}. Refer to Jacod and Shiryaev \cite[Ch. VI]{JacShir}
for background on the Skorokhod topology and weak convergence
of stochastic processes. Let $D(\R_+)$ denote the space of real valued
right-continuous functions with left-limits on $\R_+$.

\begin{Lemma}\label{lem:weak}
$X^{(n)}$ converge weakly to $X$ in the Skorokhod topology on
$D(\R_+)$, as $n\to\infty$.
\end{Lemma}

\proof.\ With regard to Corollary VII.3.6 in Jacod and Shiryaev
\cite[Ch. VI]{JacShir} it follows that it is equivalent to show
that $X^{(n)}(1)$ converges in distribution to $X(1)$, and that the
latter assertion follows if it holds that
\begin{eqnarray*}
\mu_n \to \mu, \quad \int g(u) k_n(u) \td u \to \int g(u)k(u)\td u,\\
\s^{2}_n + \int_{-1}^1 x^2k_n(x)\td x \to \s^2 + \int_{-1}^1 x^2 k(x)\td x,
\end{eqnarray*}
for any continuous bounded function $g$ that is zero around zero,
as $n\to\infty$. As a consequence of the definitions of $k_n$
and $\s^2_n$ given in \eqref{eq:kn} and \eqref{eq:sn}
it is not hard to verify that these three conditions are indeed satisfied.\exit

\begin{Lemma}\label{lem:weakXX}
$(X^{(n)}(T), \ovl {X}^{(n)}(T))$ converges in distribution to
$(X(T), \ovl X(T))$.
\end{Lemma}
\proof.\ 
It follows from Proposition VI.2.4 and remark VI.2.3 in Jacod and
Shiryaev \cite{JacShir} that for each fixed $T$ such that
$\Delta\omega(T) = 0$, the function $f:D(\R_+)\to\R $
given by
$$
f: \omega\mapsto \le(\omega(T), \sup\{\max\{\omega(s),0\}, s\leq
T\ri)
$$
is continuous (in the Skorokhod topology). In particular, if
$\omega^{(n)}\to\omega$ in the Skorokhod topology, then
$f(\omega^{(n)})\to f(\omega)$, as $n\to\infty$. In view of the fact
that L\'{e}vy processes are continuous at each fixed time a.s., it
holds that $P(\D X(T) \neq 0) = 0$, and the asserted convergence
follows in view of Lemma \ref{lem:weak}.\exit

\proof{\it of Proposition \ref{prop:convWH}.} The convergence of
$(X^{(n)}(T), \ovl {X}^{(n)}(T))$ is proved in Lemma
\ref{lem:weakXX}. If $X$ is not of type A, it is shown in Lemma
49.3 in Sato(1999) that $\ovl X(T)$ and $-\unl X(T)$ are
continuous random variables on $(0,\infty)$ (with possibly an atom
in zero if $X$ is of bounded variation). The statement
\eqref{conv1} thus follows in view of the continuity theorem.
Since $P(T(x)\leq T) = P(\unl X(T) \leq x)$, also \eqref{conv2}
follows. \exit

\end{document}